\documentclass[aps,prd,preprintnumbers,nofootinbib,superscriptaddress,twocolumn]{revtex4} 
\usepackage{ae}
\usepackage{bm} 
\usepackage{color}
\usepackage{amssymb}
\usepackage{amsmath}
\usepackage{amsfonts}
\usepackage{graphicx}
\usepackage{slashed}
\usepackage{wasysym}
\usepackage{setspace}
\usepackage{ulem}
\usepackage{multirow}
\usepackage{footnote}
\usepackage{diagbox} 
\usepackage{ulem} 

\newcommand{\stkout}[1]{\ifmmode\text{\sout{\ensuremath{#1}}}\else\sout{#1}\fi}

\newcommand{\Eqref}[1]{Eq.~\eqref{#1}}
\newcommand{\Figref}[1]{Fig.~\ref{#1}}

\allowdisplaybreaks

\begin{document}

\setlength{\unitlength}{1mm}
\title{Quantum vacuum signatures in multi-color laser pulse collisions}

\author{Holger Gies}\email{holger.gies@uni-jena.de}
\affiliation{Helmholtz-Institut Jena, Fr\"obelstieg 3, 07743 Jena, Germany}
\affiliation{GSI Helmholtzzentrum f\"ur Schwerionenforschung, Planckstra\ss e 1, 64291 Darmstadt, Germany}
\affiliation{Theoretisch-Physikalisches Institut, Abbe Center of Photonics, \\ Friedrich-Schiller-Universit\"at Jena, Max-Wien-Platz 1, 07743 Jena, Germany}
\author{Felix Karbstein}\email{felix.karbstein@uni-jena.de}
\affiliation{Helmholtz-Institut Jena, Fr\"obelstieg 3, 07743 Jena, Germany}
\affiliation{GSI Helmholtzzentrum f\"ur Schwerionenforschung, Planckstra\ss e 1, 64291 Darmstadt, Germany}
\affiliation{Theoretisch-Physikalisches Institut, Abbe Center of Photonics, \\ Friedrich-Schiller-Universit\"at Jena, Max-Wien-Platz 1, 07743 Jena, Germany}
\author{Leonhard Klar}\email{leonhard.klar@uni-jena.de}
\affiliation{Helmholtz-Institut Jena, Fr\"obelstieg 3, 07743 Jena, Germany}
\affiliation{GSI Helmholtzzentrum f\"ur Schwerionenforschung, Planckstra\ss e 1, 64291 Darmstadt, Germany}
\affiliation{Theoretisch-Physikalisches Institut, Abbe Center of Photonics, \\ Friedrich-Schiller-Universit\"at Jena, Max-Wien-Platz 1, 07743 Jena, Germany}

\date{\today}

\begin{abstract}
 Quantum vacuum fluctuations give rise to effective non-linear interactions between electromagnetic fields.
 A prominent signature of quantum vacuum nonlinearities driven by macroscopic fields are signal photons differing in characteristic properties such as frequency, propagation direction and polarization from the driving fields.
 We devise a strategy for the efficient tracing of the various vacuum-fluctuation-mediated interaction processes in order to identify the most prospective signal photon channels.
 As an example, we study the collision of up to four optical laser pulses and pay attention to sum and difference frequency generation.
 We demonstrate how this information can be used to enhance the signal photon yield in laser pulse collisions for a given total laser energy.
\end{abstract}

\maketitle

\section{Introduction}\label{sec:intro}

Although the vacuum is the state characterized by the absence of real particles it constitutes a portal to the particle degrees of freedom of the underlying quantum field theory due to the omnipresence of quantum fluctuations.
In quantum electrodynamics (QED), which governs the interaction of light and matter within the Standard Model of particle physics, these fluctuations involve virtual electrons, positrons and photons.
Electromagnetic fields provide a promising means to probe these quantum vacuum 
fluctuations: as electromagnetic fields couple to charges, vacuum fluctuations 
involving charged particles mediate effective interactions between them 
\cite{Euler:1935zz,Heisenberg:1935qt,Weisskopf:1936kd}. The latter supplement 
Maxwell's classical theory of electrodynamics with nonlinear self-couplings of 
the electromagnetic field. Using Heaviside-Lorentz units with $c=\hbar=1$, 
the metric convention $g_{\mu\nu}=\mathrm{diag}(-,+,+,+)$,
these quantum vacuum nonlinearities are parametrically suppressed by inverse powers of
$m_e^2/e\simeq1.3\times10^{18}\,\frac{\rm V}{\rm m}\simeq4.4\times10^9\,{\rm T}$. Here $e$ is the elementary charge and $m_e$ the electron mass, setting the reference scale the applied electromagnetic fields are compared to.
Due to the fact that the strongest macroscopic fields available in the laboratory fulfill $E\simeq{\cal O}(10^{14})\frac{\rm V}{\rm m}$ and $B\simeq{\cal O}(10^6){\rm T}$, the induced interactions are generically very small and elusive in experiment.

All-optical probes provide a prominent route toward verifying QED vacuum nonlinearities in a controlled laboratory experiment with macroscopic electromagnetic fields.
Various theoretical proposals studied in the literature assume these electromagnetic fields to be delivered by high-intensity lasers; see the reviews~\cite{Marklund:2006my,DiPiazza:2011tq,Battesti:2012hf,King:2015tba,Hill:2017,Inada:2017lop,Karbstein:2019oej,Schoeffel:2020svx} and references therein. 
The basic idea is to look for signal photons which are induced in the effective interaction of several driving laser fields and differ in key properties, such as polarization, frequency and propagation direction, from the photons constituting the latter.
However, separating the typically small signal from the large background in general constitutes a major challenge \cite{Tommasini:2010fb,King:2013am,Karbstein:2019dxo,Karbstein:2020gzg}.  
Aiming at a systematic enhancement of photonic quantum vacuum signals, a detailed knowledge about the microscopic origin of the prospective signal photon channels is indispensable.
In this article, we demonstrate how their microscopic origin can be efficiently traced.
Moreover, using a particular scenario envisioning the collision of several high-intensity laser pulses as an example, we show how this information can be used to enhance the signal photon yield for a given total laser energy.

Our article is organized as follows: in Sec.~\ref{sec:form} we briefly recall the vacuum emission picture used in the present article to study all-optical signatures of quantum vacuum nonlinearity in macroscopic electromagnetic fields as provided by high-intensity lasers.
Moreover, we introduce several definitions relevant for the subsequent discussion.
Thereafter, in Sec.~\ref{sec:scen} we apply the formalism outlined in Sec.~\ref{sec:form} to an exemplary experimental scenario based on the collision of several high-intensity laser pulses.
Following more general considerations about a prospective collision geometry involving beams of several colors, we analyze and outline the frequency and directional characteristics of the attainable signals and assess the possibility of their measurement against the background of the driving laser photons.
In Sec.~\ref{sec:channelanalysis} we demonstrate in detail how such studies can be substantially enhanced and simplified.
In particular, interpretations in terms of elastic and inelastic scattering processes turn out to be useful.
These considerations highlight an aspect of the vacuum emission approach which was not yet fully exploited in previous all-optical quantum vacuum studies.
Namely, we show how the microscopic origin of the signal can be efficiently traced and analyzed. Moreover, we sketch how the information extracted along these lines can be employed to enhance prospective signals. 
Finally, we end with conclusions and a short outlook in Sec.~\ref{sec:conc}.

\section{Formalism}\label{sec:form}

All-optical signatures of quantum vacuum nonlinearity can be efficiently 
analyzed in terms of vacuum emission processes 
\cite{Galtsov:1971xm,Karbstein:2014fva}: in the interaction region where the 
strong driving laser fields overlap signal photons are generated. These signal 
photons are to be detected far outside the interaction region and constitute 
the signature of quantum vacuum nonlinearity in experiment.
For state-of-the-art high-intensity laser fields of optical and near-infrared 
frequencies $\omega\ll m_e$ reaching electric and magnetic peak field strengths 
$\{E,B\}\ll\frac{e}{m_e^2}$, such a study can be based on the leading 
contribution to the one-loop Heisenberg-Euler effective Lagrangian ${\cal 
L}_{\rm HE}$  
\cite{Euler:1935zz,Heisenberg:1935qt,Weisskopf:1936kd,Karplus:1950zz,Schwinger:1951nm,Ritus:1975cf,Dittrich:1985yb,Dittrich:2000zu,Dunne:2004nc,Gies:2016yaa}.
The latter is a function of  the two scalar invariants of the electromagnetic 
field,
\begin{align}
\mathcal{F}&=\frac{1}{4} F^{\mu\nu} F_{\mu\nu}= \frac{1}{2}\left(\mathbf{B}^2-{\bf E}^2\right)\,, \nonumber \\
\mathcal{G}&=\frac{1}{4} \tilde{F}^{\mu\nu}F_{\mu\nu}= -\mathbf{B}\cdot{\bf E}\,, \label{eq:invariantFG}
\end{align}
with the metric convention $g_{\mu\nu}=\mathrm{diag}(-,+,+,+)$,
and can be decomposed as ${\cal L}_{\rm HE}={\cal L}_{\rm MW}+{\cal L}_{\rm int}$.
Here, ${\cal L}_{\rm MW}=-{\cal F}$ is the classical Maxwell Lagrangian and 
\begin{equation}
\frac{\mathcal{L}_{\text{int}}}{m_e^4}= \frac{1}{360\pi^2} \Bigl(\frac{e}{m_e^2} \Bigr)^4\left(4 \mathcal{F}^2 + 7 \mathcal{G}^2 \right) + \mathcal{O}\!\left(\big(\tfrac{eF}{m_e^2}\bigr)^6\right) \label{eq:Leff}
\end{equation}
encodes the effective nonlinear interactions of the electromagnetic field induced by QED vacuum fluctuations.
As detailed in Refs.~\cite{Karbstein:2014fva,Gies:2017ygp,Karbstein:2019oej}, the differential number of signal photons ${\rm d}^3N_{\left(p\right)}$ of polarization $p$ which have an energy ${\rm k}=|\bf{k}|$ in the differential energy interval ${\rm dk}$ and are emitted into the solid angle ${\rm d}\Omega$ 
around $\hat{{\bf k}}$ follows from the zero-to-single signal photon transition amplitude ${\cal S}_{(p)}({\bf k})$ as
\begin{equation}
{\rm d}^3N_{(p)}=\frac{{\rm k^2}{\rm dk}\,{\rm d}\Omega}{(2\pi)^3}\,|{\cal S}_{(p)}({\bf k})|^2\,.\label{eq:d3Nsig}
\end{equation}
The signal photon amplitude can be determined from $\Gamma_{\rm int}[\hat{a}(x)]=\int{\rm d}^4x\,{\cal L}_{\rm int}|_{F\to F+\hat{f}}$ upon splitting the electromagnetic field as $F^{\mu\nu}\to F^{\mu\nu}+\hat{f}^{\mu\nu}$ into a classical background field $F^{\mu\nu}$ and a operator-valued signal photon field $\hat{a}^\mu$, with field-strength tensor $\hat{f}^{\mu\nu}=\partial^\mu\hat{a}^\nu-\partial^\nu\hat{a}^\mu$.
It is given by
\begin{align}
S_{(p)}\left(\mathbf{k}\right) &= \bigl\langle\gamma_{\left(p\right)}\left(\mathbf{k}\right)\bigr|\Gamma_{\text{int}}[\hat{a}(x)]\bigl|0\bigr\rangle\nonumber \\
& = \mathrm{i}\frac{\epsilon^{*\mu}_{\left(p\right)}\bigl(\hat{\bf k}\bigr)}{\sqrt{2k^0}} \int \mathrm{d}^4x\; \mathrm{e}^{\mathrm{i}k_{\alpha}x^{\alpha}} \nonumber \\
& \phantom{=} \times  \Bigl( k^{\nu} F_{\nu\mu} \frac{\partial \mathcal{L}_{\text{int}}}{\partial \mathcal{F}}  + k^{\nu} \tilde{F}_{\nu\mu} \frac{\partial \mathcal{L}_{\text{int}}}{\partial \mathcal{G}} \Bigr)\biggl|_{k^0={\rm k}}\,, \label{eq:S_p_Gamma}
\end{align}
where $\epsilon^{\mu}_{\left(p\right)}\bigl(\hat{\bf k}\bigr)=(0,{\bf e}_{(p)})$ is the polarization vector of the induced signal photon; $*$ denotes complex conjugation.
We span the signal photon polarizations by two transverse vectors ${\bf e}_{(p)}$ with $p\in\{1,2\}$, fulfilling $\hat{\bf k}\times{\bf e}_{(p)}={\bf e}_{(p+1)}$ and ${\bf e}_{(3)}=-{\bf e}_{(1)}$.
The derivatives for ${\cal F}$ and ${\cal G}$ entering \Eqref{eq:S_p_Gamma} follow straightforwardly from \Eqref{eq:Leff},
\begin{equation}
\left\{\begin{array}{c}
\!\!\frac{\partial \mathcal{L}_{\text{int}}}{\partial \mathcal{F}}\!\! \\
\!\!\frac{\partial \mathcal{L}_{\text{int}}}{\partial \mathcal{G}}\!\!
\end{array}\right\} = \frac{1}{45}\frac{e^2}{4\pi^2} \Bigl(\frac{e}{m_e^2}\Bigr)^2 \left\{\begin{array}{c}
\!4 \mathcal{F}\left(x\right)\!\! \\
\!7 \mathcal{G}\left(x\right)\!\!
\end{array}\right\} + \mathcal{O}\left(\bigl(\tfrac{e \bar{F}}{m_e^2}\bigr)^4\right).
\end{equation}
Upon plugging these quantities into \Eqref{eq:S_p_Gamma} and limiting ourselves to the terms written out explicitly, we obtain 
\begin{align}
S_{(p)} \left(\mathbf{k}\right) &= \frac{1}{\mathrm{i}} \frac{e}{4\pi^2}  \frac{m_e^2}{45} \sqrt{\frac{\rm k}{2}} \left(\frac{e}{m_e^2}\right)^3 \int \mathrm{d}^4x\; \mathrm{e}^{\mathrm{ik}(\hat{\bf k}-t)} \nonumber \\
&\phantom{=} \times \Bigl( 4\left[\mathbf{e}_{\left(p\right)}\cdot{\bf E} - \mathbf{e}_{\left(p+1\right)}\cdot \mathbf{B}\right] \mathcal{F} \nonumber\\
&\phantom{= \times \Bigl(} +7 \left[\mathbf{e}_{\left(p\right)}\cdot\mathbf{B} + \mathbf{e}_{\left(p+1\right)}\cdot {\bf E}\right] \mathcal{G} \Bigr)\,. \label{eq:S1_FG}
\end{align}

In this article, we focus on the collision of $n+1$ linearly polarized paraxial laser fields characterized by the electric and magnetic field vectors ${\bf E}_i={\cal E}_i(x)\,\hat{\bf E}_i$ and ${\bf B}_i={\cal E}_i(x)\,\hat{\bf B}_i$, with $i\in\{0,1,\ldots,n\}$, fulfilling $\hat{\bf E}_i\cdot\hat{\bf B}_i=0$.
The associated unit wave vectors are $\hat{\pmb{\kappa}}_i=\hat{\bf E}_i\times\hat{\bf B}_i$.
In this case, the photon transition amplitude~\eqref{eq:S1_FG} can be expressed as $S_{\left(p\right)} \left(\mathbf{k}\right)=\sum_{i,j,l} S_{\left(p\right);ijl} \left(\mathbf{k}\right)$, with
\begin{equation}
S_{\left(p\right);ijl} \left(\mathbf{k}\right) = \frac{1}{\mathrm{i}} \frac{e}{4\pi^2}\frac{m_e^2}{45} \sqrt{\frac{\rm k}{2}}\, \Bigl(\frac{e}{m_e^2}\Bigr)^3  \mathcal{I}_{ijl} \left(\mathbf{k}\right) g_{\left(p\right);ijl}\bigl(\hat{\mathbf{k}}\bigr). \label{eq:S_p}
\end{equation}
Here, the entire dependence on the field profiles ${\cal E}_i(x)$ is encoded in the quantity
\begin{equation}
 \mathcal{I}_{ijl}\left(\mathbf{k}\right) = \int\!\mathrm{d}^4x\, \mathrm{e}^{\mathrm{ik}(\hat{\bf k}\cdot\bf x-t)} \mathcal{E}_i(x)\mathcal{E}_j(x)\mathcal{E}_l(x)\,, \label{eq:FourierInt}
\end{equation}
and the dependence on the polarization assignments of the beams and the collision geometry in
\begin{align}
g_{(p);ijl}(\hat{\bf k}\bigr) &= 2 \bigl( \mathbf{e}_{(p)} \!\cdot \hat{\bf E}_l  - \mathbf{e}_{(p+1)} \!\cdot \hat{\bf B}_l \bigr) \bigl( \hat{\bf B}_i \cdot \hat{\bf B}_j  - \hat{\bf E}_i \cdot \hat{\bf E}_j \bigr) \nonumber\\
& - \frac{7}{2} \big( \mathbf{e}_{(p)} \!\cdot \hat{\bf B}_l  + \mathbf{e}_{(p+1)} \!\cdot \hat{\bf E}_l \bigr)  \bigl(\hat{\bf B}_i \cdot \hat{\bf E}_j + \hat{\bf B}_j \cdot \hat{\bf E}_i \bigr). \label{eq:geo_func}
\end{align}

In the following, we parameterize the emission directions of the signal photons as $\hat{\bf k}=(\cos\varphi\sin\vartheta,\sin\varphi\sin\vartheta,\cos\vartheta)$, such that ${\rm d}^2\Omega={\rm d}\varphi\,{\rm d}\!\cos\vartheta$.
We choose the vectors spanning the polarization basis of the signal photons as $\mathbf{e}_{(1)}(\beta)=\hat{\bf k}|_{\vartheta\to\vartheta+\frac{\pi}{2}}\cos\beta+\hat{\bf k}|_{\vartheta=\frac{\pi}{2},\varphi\to\varphi+\frac{\pi}{2}}\sin\beta$
and $\mathbf{e}_{(2)}(\beta)=\mathbf{e}_{(1)}(\beta+\pi/2)$, where $\beta$ is an {\it a priori} arbitrary angle; its choice fixes a specific polarization basis.
Moreover, we introduce the number density $\rho_{(p)}(\varphi,\vartheta|{\rm k}_{\rm min},{\rm k}_{\rm max})$ of signal photons of energies $\rm k$ constrained by ${\rm k}_{\rm min}\leq{\rm k}\leq{\rm k}_{\rm max}$. 
The latter is obtained from \Eqref{eq:d3Nsig} upon integration over energy as
\begin{equation}
\rho_{(p)}(\varphi,\vartheta|{\rm k}_{\rm min},{\rm k}_{\rm max}) = \frac{1}{(2\pi)^3} \!\int_{\mathrm{k}_{\rm min}}^{\mathrm{k}_{\rm max}} \!\!\!\mathrm{dk}\,{\rm k}^2\bigl|S_{(p)}(\mathbf{k})\bigr|^2.
\label{eq:rho_result}
\end{equation}
In the present case, the modulus squared of the signal photon amplitude entering \Eqref{eq:rho_result} can be expressed as
\begin{equation}
 \bigl|S_{(p)}(\mathbf{k})\bigr|^2
 =\sum_{\ell,\ell'}
 {\rm Re}\bigl\{S_{(p);\ell}(\mathbf{k})S^*_{(p);\ell'}(\mathbf{k})\bigr\}\,,\label{eq:SpSpstar}
\end{equation}
where the sum runs over all sets $\ell=\{i,j,l\}$ and $\ell'=\{i',j',l'\}$.
From \Eqref{eq:rho_result} the number of signal photons of polarization $p$ and energies ${\rm k}_{\rm min}\leq{\rm k}\leq{\rm k}_{\rm max}$ emitted into the solid angle ${\cal A}$ follows as
\begin{equation}
 N_{(p)}({\cal A}|{\rm k}_{\rm min},{\rm k}_{\rm max})=\int_{\cal A}{\rm d}\Omega\,\rho_{(p)}(\varphi,\vartheta|{\rm k}_{\rm min},{\rm k}_{\rm max})\,. \label{eq:NpA}
\end{equation}
Also note that the signal photon density and number accessible in a polarization insensitive measurement follow upon summation over the two transverse polarizations $p\in\{1,2\}$. They are given by $\rho(\varphi,\vartheta|{\rm k}_{\rm min},{\rm k}_{\rm max})= \sum_{p=1}^2\rho_{(p)}(\varphi,\vartheta|{\rm k}_{\rm min},{\rm k}_{\rm max})$ and $N({\cal A}|{\rm k}_{\rm min},{\rm k}_{\rm max})=\sum_{p=1}^2N_{(p)}({\cal A}|{\rm k}_{\rm min},{\rm k}_{\rm max})$, respectively. Obviously, the total number of emitted signal photons is $N_{\rm tot}=N(4\pi|0,\infty)$.

\section{Example Scenario}\label{sec:scen}

Let us apply the approach devised in Sec.~\ref{sec:form} to a specific, experimentally viable scenario involving the collision of several high-intensity laser pulses.
Our main focus is on the analysis and reconstruction of properties of the microscopic scattering processes giving rise to the dominant signal photon emission channels.
Special attention is paid to signal photon contributions which allow for a clear signal-to-background separation in experiment. Prominent criteria facilitating such a separation are, e.g., a distinct emission direction outside the forward cones of the driving beams, or a frequency outside their spectra allowing for an unobstructed detection of the signal.

\subsection{Collision geometry and beam model}
\label{sec:beammodel}

\begin{figure}[t]
	\centering
	\includegraphics[width=0.9\columnwidth]{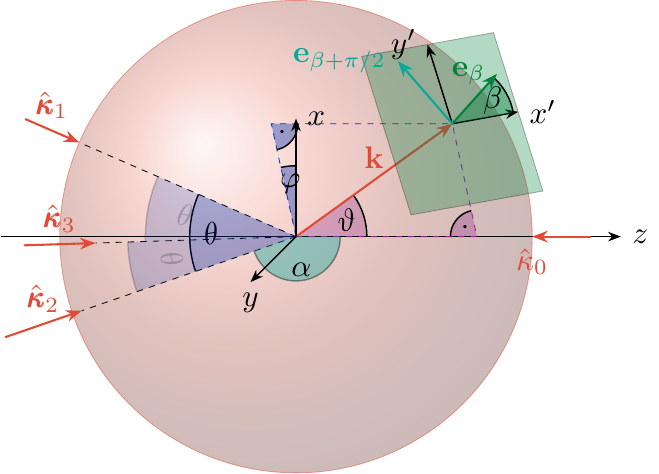}
	\caption{Illustration of the scenario considered in the main text: three high-intensity pulses of unit wave vectors $\hat{\pmb{\kappa}}_i$, with $i\in\{1,2,3\}$, are superposed to create a strongly peaked field configuration in their overlap region; each pair of these beams encloses the same angle $\theta$. An additional high-intensity laser pulse with $\hat{\pmb{\kappa}}_0$ collides with the apex of the pyramid formed by beams $1$-$3$ under an angle of $\alpha$. The wave vector of the signal photons is denoted by $\mathbf{k}$. It encodes the signal frequency in its length (radius of the red sphere), and the emission direction in its orientation parameterized by the angles $\vartheta$ and $\varphi$. The possible polarizations of the signal are parameterized by the angle $\beta$ spanning a plane tangential to the sphere.}
	\label{fig:sphere_angle}
\end{figure}

For definiteness, we use a collision geometry involving $n+1$ driving laser pulses \cite{Klar:2020ych}: beam $0$ collides with the apex of the regular pyramid formed by the beam axes of $n$ additional laser beams; see Fig.~\ref{fig:sphere_angle} for an illustration.
Here, we focus on a scenario involving four driving laser fields, i.e., $n=3$. 
These are envisioned to be generated by a single high-intensity laser system of the ten petawatt class, such as available at the Extreme Light Infrastructure Nuclear Physics (ELI-NP) project \cite{Lureau:2020,ELI-NP}, by employing beam-splitting and frequency doubling techniques. More specifically, we assume the initial laser system to deliver pulses of energy $W=250\,{\rm J}$ and duration $\tau=25\,{\rm fs}$ at a wavelength of $\lambda=800\,{\rm nm}$.
While each of the four laser fields generated in this way features exactly this pulse duration, we assume them to have different frequencies: beams $0$ and $1$ are fundamental frequency beams with $\omega_0=\frac{2\pi}{\lambda}\simeq1.55\,\text{eV}$, beam $2$ is frequency-doubled, and beam $3$ is frequency-quadrupled.

Moreover, we assume that each pair of the beams $1$-$3$ encloses the same angle $\theta$, such that
 \begin{equation}
 \hat{\pmb{\kappa}}_i \cdot \hat{\pmb{\kappa}}_j = (1-\cos\theta)\delta_{ij} + \cos\theta
 \end{equation}
for $i,j\in\{1,2,3\}$ with Kronecker delta $\delta_{ij}$. Accordingly we choose these unit wave vectors as
\begin{align}
\hat{\pmb{\kappa}}_i = 
\left(\begin{array}{c}
	-a\cos\bigl(2\pi\frac{i-1}{3}\bigr)\vspace{1mm} \\
	-a\sin\bigl(2\pi\frac{i-1}{3}\bigr) \vspace{1mm} \\
	\sqrt{1-a^2}
\end{array}
\right),
\end{align}
with
$a=\sqrt{2(1-\cos{\theta})/3}$.
On the other hand, the wave vector of the beam $0$ colliding with the apex of the pyramid formed by beams $1$-$3$ is
$ \hat{\pmb{\kappa}}_0=-\hat{\mathbf{e}}_z$.
The angle between this and any other beam is $\alpha=\pi-\arctan(a/\sqrt{1-a^2})$.
In the remainder of this work, we set $\theta=\frac{\pi}{2}=90^{\circ}$, such that $\alpha=\pi-\arctan\sqrt{2}\approx 125.26^{\circ}$.

We assume the driving laser fields to be well-described as paraxial Gaussian beams.
In order to facilitate a concise analytical analysis of the expected signals, we resort to an {\it infinite Rayleigh range approximation}. This approximation neglects the widening of the laser pulses as a function of the longitudinal coordinate ${\bf x}\cdot\hat{\pmb{\kappa}}_i$ measured from the beam focus at ${\bf x}={\bf 0}$, where the temporal pulse envelope reaches its maximum at $t=0$.
This is a reasonable approximation of the driving laser fields in the vicinity of their beam foci.
We thus expect our predictions for the attainable signal photon numbers to be quantitatively accurate in the specific collision scenario considered here, where the collision angles between each pair of laser beams $\theta_{i,j}=\varangle(\hat{\pmb{\kappa}}_i,\hat{\pmb{\kappa}}_j)$ fulfill $0^\circ\leq \varangle(\hat{\pmb{\kappa}}_i,\hat{\pmb{\kappa}}_j)\leq130^\circ$ \cite{Gies:2017ygp}.
Note that, in principle, this analysis could also be performed directly with paraxial Gaussian beams \cite{Gies:2017ygp} or even generic laser fields fulfilling Maxwell's equations in vacuo exactly \cite{Blinne:2018nbd}
even though this is technically or numerically more demanding.
Since all driving laser pulses are assumed to have the same pulse duration $\tau$, 
the field profile of the $i$th laser beam can be approximated in the interaction region as \cite{Gies:2017ygp}
\begin{equation}
\mathcal{E}_i \left(x\right) =  \mathfrak{E}_{i}\, \mathrm{e}^{-(\frac{{\bf x}\cdot\hat{\pmb{\kappa}}_i-t}{\tau/2})^2}\, \mathrm{e}^{- \frac{{\bf x}^2-({\bf x}\cdot\hat{\pmb{\kappa}}_i)^2}{w_{i}^2}}  \cos\bigl(\omega_i({\bf x}\cdot\hat{\pmb{\kappa}}_i-t)\bigr)\,,  \label{eq:E_x_infRay}
\end{equation}
where $\mathfrak{E}_i$, $w_i$ and $\omega_i$ are the peak field amplitude, beam waist and oscillation frequency, respectively. Throughout this work, we consider only optimal laser pulse collisions, i.e., all laser beams are focused on the same spot and reach their peak field values at the same time.

Besides, we fix the linear polarizations of the laser fields, by choosing $\hat{\mathbf{E}}_0=\mathbf{e}_y$ and
$\mathbf{e}_y\cdot\hat{\mathbf{E}}_i=0$ for $i\in\{1,2,3\}$.
This choice is motivated by the observation
that the total number of signal photons attainable in a polarization insensitive measurement is maximized for counter-propagating beams with a relative polarization difference of $\pi/2$ \cite{Karbstein:2019bhp}.
We have explicitly checked that this is also the case for our setup, where beam 0 can be considered as effectively counter-propagating the combined field of beams 1-3.
Together with the transversality condition $\hat{\pmb{\kappa}}_i\cdot\hat{\mathbf{E}}_i=0$ the above choice determines the polarization vectors of all laser fields up to a sign.
Ensuring a positive sign for the $\rm x$ component, the polarization vectors of beams $i\in\{1,2,3\}$ read
\begin{align}
 \hat{\mathbf{E}}_i= \frac{1}{\sqrt{1-a^2\sin^2(2\pi\frac{i-1}{3})}}
\left( \begin{array}{c}
	\sqrt{1-a^2} \\
	0 \\
 a\cos(2\pi\frac{i-1}{3})
 \end{array}
 \right).
\end{align}
The unit vectors for the associated magnetic fields are $\hat{\mathbf{B}}_i=\hat{\pmb{\kappa}}_i\times\hat{\mathbf{E}}_i$.

\subsection{Beam splitting and losses} \label{sec:amplitudes_losses}

The peak field amplitude of a laser beam of pulse energy $W$ and duration $\tau$ focused to a waist spot size of $w_0$ is \cite{Karbstein:2017jgh,Karbstein:2019oej}
\begin{equation}
 \mathcal{E}_{\star} = \sqrt{8\sqrt{\frac{2}{\pi}} \frac{W}{\pi w_0^2 \tau}} \label{eq:E_star}\,.
\end{equation}
For a diffraction-limited Gaussian beam of wavelength $\lambda$, we have $w_0\simeq\lambda$.
Throughout this work, we assume all the individual laser fields to be focused to the same waist spot size $w_i=w_0=\lambda$ and measure their peak field amplitudes in units of the peak field ${\cal E}_{\star}$ which could be achieved by focusing the initial laser pulse of energy $W=250\,{\rm J}$ to its diffraction limit.
In turn, we have $\mathfrak{E}_i=A_i{\cal E}_\star$, where $A_i$ denote dimensionless amplitudes, also accounting for potential losses.
Analogously, we measure the oscillation frequencies of the beams in units of $\omega_0$, such that $\omega_i=\nu_i\omega_0$, with dimensionless amplitude $\nu_i$. In the present case, here we have $\nu_0=\nu_1=1$, $\nu_2=2$ and $\nu_3=4$. 

\begin{figure}[t]
	\centering
	\includegraphics[width=0.8\columnwidth]{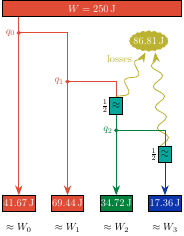}
	\caption{Schematic of the beam splitting and higher harmonic generation processes invoked to create four beams $i\in\{0,1,2,3\}$ of pulse energy $W_i$ from a single high-intensity laser system delivering pulses of energy $W=250\,\text{J}$ and frequency $\omega_0=1.55\,\text{eV}$. The partitioning proceeds in three stages, with the associated partition factors given by $q_0$, $q_1$ and $q_2$.
Beams $2$ and $3$ are frequency-doubled and quadrupled, respectively. Each frequency doubling comes with a loss of $50\%$. We depict the scenario with $q_0=5/6$. The values of $q_1$ and $q_2$ are chosen such that $W_1=2W_2=4W_3$.}
	\label{fig:Amplitude_Division}
\end{figure}

\begin{figure}[t]
	\centering
	\includegraphics[scale=0.45]{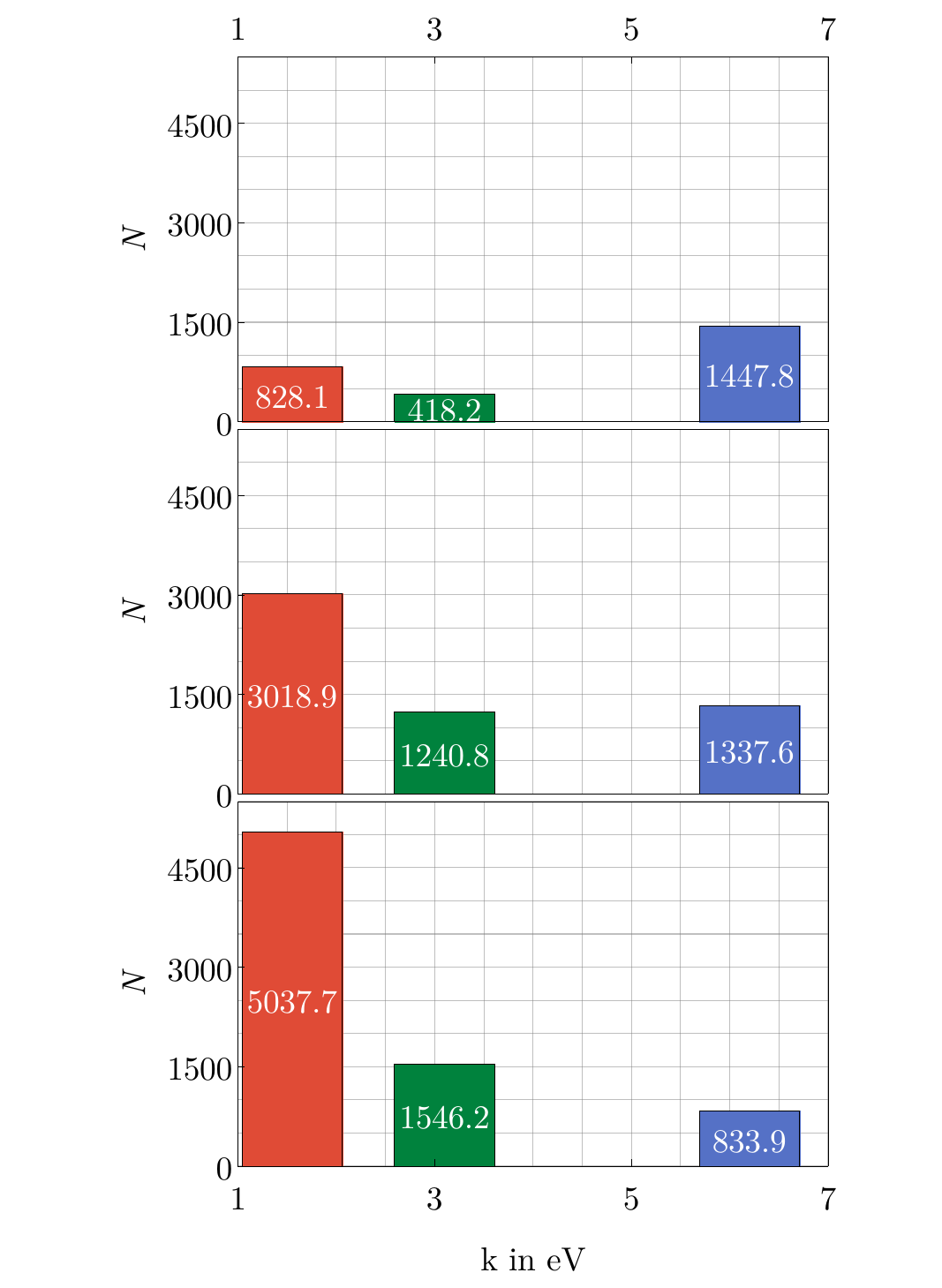}
	\caption{Signal photon spectra associated with the three different pulse-energy distributions of the driving laser fields (a)-(c) (top to bottom) listed in Tab.~\ref{tab:q_values}.  For each distribution we provide the values of $N(4\pi|{\rm k_{min}},{\rm k_{max}})$ for the frequency regimes $1.04\,\mathrm{eV}\lesssim{\rm k}\lesssim2.06\,\mathrm{eV}$ (red), $2.59\,\mathrm{eV} \lesssim{\rm k}\lesssim 3.61\,\mathrm{eV}$ (green) and $5.69\,\mathrm{eV} \lesssim{\rm k}\lesssim 6.71\,\mathrm{eV}$ (blue). The respective signal photon number is written inside the particular bar.}
	\label{fig:NlA}
\end{figure}

Each frequency doubling process comes with a loss: we conservatively estimate the energy loss for the conversion process preserving the pulse duration as $50\%$ \cite{Marcinkevicius:2004}.
Correspondingly, the energies $W_i=A_i^2 W$ of all beams do not add up to $W$ but to $W^{\rm eff}=\sum_{i=0}^3W_i< W$.
Only, for vanishing losses we would have $W^{\rm eff}=W$.
Here, we assume the beam splitting and higher harmonic generation to proceed in several steps. First, the original laser pulse of energy $W$ is split into two parts: the part with energy $W_0=(1-q_0)W$ constitutes beam $0$, and the remainder of energy $q_0W$ is to be subdivided further; the factor $0<q_0<1$ controls the partitioning ratio.
Second, the remaining energy $q_0W$ is again partitioned into a fundamental frequency part of energy $W_1=q_0(1-q_1)W$ constituting beam $1$, and another one of energy $q_0q_1W$ which undergoes frequency doubling; as above $0<q_1<1$.
Accounting for the loss of $50\%$ associated with the frequency doubling process, the latter contribution results in a frequency-doubled pulse of energy $\frac{1}{2}q_0q_1W$.
In the last step, the procedure is repeated for the frequency-doubled pulse with a partitioning factor of $0<q_2<1$. This results in an energy of $W_2=\frac{1}{2}q_0q_1(1-q_2)W$ for the frequency-doubled beam $2$ and an energy of $W_3=\frac{1}{4}q_0q_1q_2W$ for the frequency-quadrupled beam $3$.
Hence, in the present scenario we have $A_0=\sqrt{1-q_0}$, $A_1=\sqrt{q_0(1-q_1)}$, $A_2=\sqrt{q_0q_1(1-q_2)/2}$ and $A_3=\sqrt{q_0q_1q_2}/2$.
Figure~\ref{fig:Amplitude_Division} illustrates this procedure for a specific choice of the partition factors; see Fig.~\ref{fig:NlA} below for the influence of different choices of the partition factors on the signal photon numbers.

So far, we did not specify a particular choice of the dimensionless field amplitudes $A_i$ of the driving laser beams. As detailed above, in an experiment these amplitudes can be adjusted by choosing the partition factors $q_0$, $q_1$ and $q_2$ accordingly.
For definiteness, we choose $q_0=\frac{5}{6}$ in the following. 
In this way, a substantial fraction of thetotal laser energy is put into the beam that collides with theapex of the pyramid formed by the other beams.

We limit our discussion to three example distributions of the pulse energies of beams $1$-$3$ forming the pyramid: either the pulse energy of each higher frequency component is doubled (such that $W_1:W_2:W_3=1:2:4$), bisected ($W_1:W_2:W_3=4:2:1$), or
quartered ($W_1:W_2:W_3=16:4:1$).
The explicit values of the required partition factors $q_1$ and $q_2$ are listed in Tab.~\ref{tab:q_values}, together with the respective effective energy $W^{\rm eff}$ put into the interaction region by all four driving laser pulses. Correspondingly, the associated energy loss is given by $W^{\rm loss}=W-W^{\rm eff}$.

\begin{table}[t]
	\centering
	\caption{Examples of different pulse-energy distributions for beams $1$-$3$. For each choice, we provide the values of the partition factors $q_1$ and $q_2$ required to ensure a given distribution for fixed $q_0=5/6$. $W^{\rm eff}$ is the effective energy put into the interaction region by all four beams. The initially available total laser energy is $W=250\,{\rm J}$, such that the associated energy loss is $W^{\rm loss}=W-W^{\rm eff}$.}
	\label{tab:q_values}
	\begin{tabular}{p{0.05\columnwidth}<{\centering}p{0.25\columnwidth}<{\centering}p{0.15\columnwidth}<{\centering}p{0.15\columnwidth}<{\centering}p{0.15\columnwidth}<{\centering}p{0.15\columnwidth}<{\centering}p{0.15\columnwidth}<{\centering}}
	\hline \hline
	&$W_1:W_2:W_3$ & $q_1$ & $q_2$ & $W^{\rm eff}[{\rm J}]$ & $W^{\rm loss} [{\rm J}] $ \\ \hline
	(a)& $1:2:4$ & $\;20/21\;$ & $4/5$ & 111.11 & 138.89 \\
	(b)& $4:2:1$ & $2/3$ & $1/2$ & 163.19 & 86.81 \\
	(c)& $16:4:1$ & $3/7$ & $1/3$ & 197.92 & 52.08 \\
	\hline \hline
	\end{tabular}
\end{table}

Equation~\eqref{eq:NpA} allows to determine the number of signal photons $N(4\pi|{\rm k_{min}},{\rm k_{max}})$ for each distribution.
In the present scenario, we find substantial signal photon contributions in the three distinct frequency regimes $1.04\,\mathrm{eV}\lesssim{\rm k}\lesssim2.06\,\mathrm{eV}$, $2.59 \,\mathrm{eV}\lesssim{\rm k}\lesssim 3.61\,\mathrm{eV}$ and $5.69\,\mathrm{eV} \,\mathrm{eV}\lesssim{\rm k}\lesssim 6.71\,\mathrm{eV}$, centered at the frequencies of the driving laser beams $\omega_0$, $2\omega_0$ and $4\omega_0$, respectively.
The width of each of these regimes is $1.02\,{\rm eV}$ and has been chosen such as to cover the full signal; cf. also the discussion in Sec.~\ref{sec:distribution} below.
See Fig.~\ref{fig:Amplitude_Division} for the signal photon numbers associated with these frequency regimes for the different pulse-energy distributions considered in Tab.~\ref{tab:q_values}: in the first approach we double the pulse energy put into each higher harmonic.
This leads to a relatively high number of signal photons with ${\rm k}\simeq4\omega_0$ as compared to the other regimes.
Interestingly, even when bisecting the pulse energy put into each higher harmonic, the number of signal photons with ${\rm k}\simeq4\omega_0$ slightly surpasses that for ${\rm k}\simeq2\omega_0$. However, most signal photons are induced at ${\rm k}\simeq\omega_0$.
When quartering the energy put into each higher harmonic most signal photons are again found at ${\rm k}\simeq\omega_0$, but this time the amount of signal photons with ${\rm k}\simeq4\omega_0$ is smaller than that for ${\rm k}\simeq2\omega_0$.
As is obvious from Tab.~\ref{tab:q_values}, higher pulse energies of the frequency doubled and quartered beams imply larger losses.

Apart from these signals, it is noteworthy that we find a clear signal in the frequency regime $7.24\,\mathrm{eV} \lesssim{\rm k}\lesssim 8.26\,{\rm eV}$ peaked at a frequency of $5\omega_0$.
However, due to the substantially smaller amount of signal photons associated with this frequency regime, we do not display this signal in Fig.~\ref{fig:NlA}.
For the pulse-energy distribution (a) we count $1.07$, for (b) $2.81$, and for (c) $1.64$ signal photons per shot in this frequency regime.
On the other hand, the fact that this signal lies outside the frequencies of the driving laser fields implies the possibility of an essentially background-free detection.

The fact that the clearly discernible signal at $5\omega_0$ becomes maximum for the pulse-energy distribution (b) with $q_1=2/3$ and $q_2=1/2$
motivates us to focus on this choice in the remainder of this article.
For completeness, we note that the relative amplitudes associated with this choice are $A_0=1/\sqrt{6}$, $A_1=\sqrt{5/2}/3$, $A_2= \sqrt{5}/6$ and $A_3=\sqrt{5/2}/6$.

\subsection{Frequency and directional characteristics of the signal} \label{sec:distribution}

\begin{figure*}[t]
	\centering
	\includegraphics[width=\textwidth]{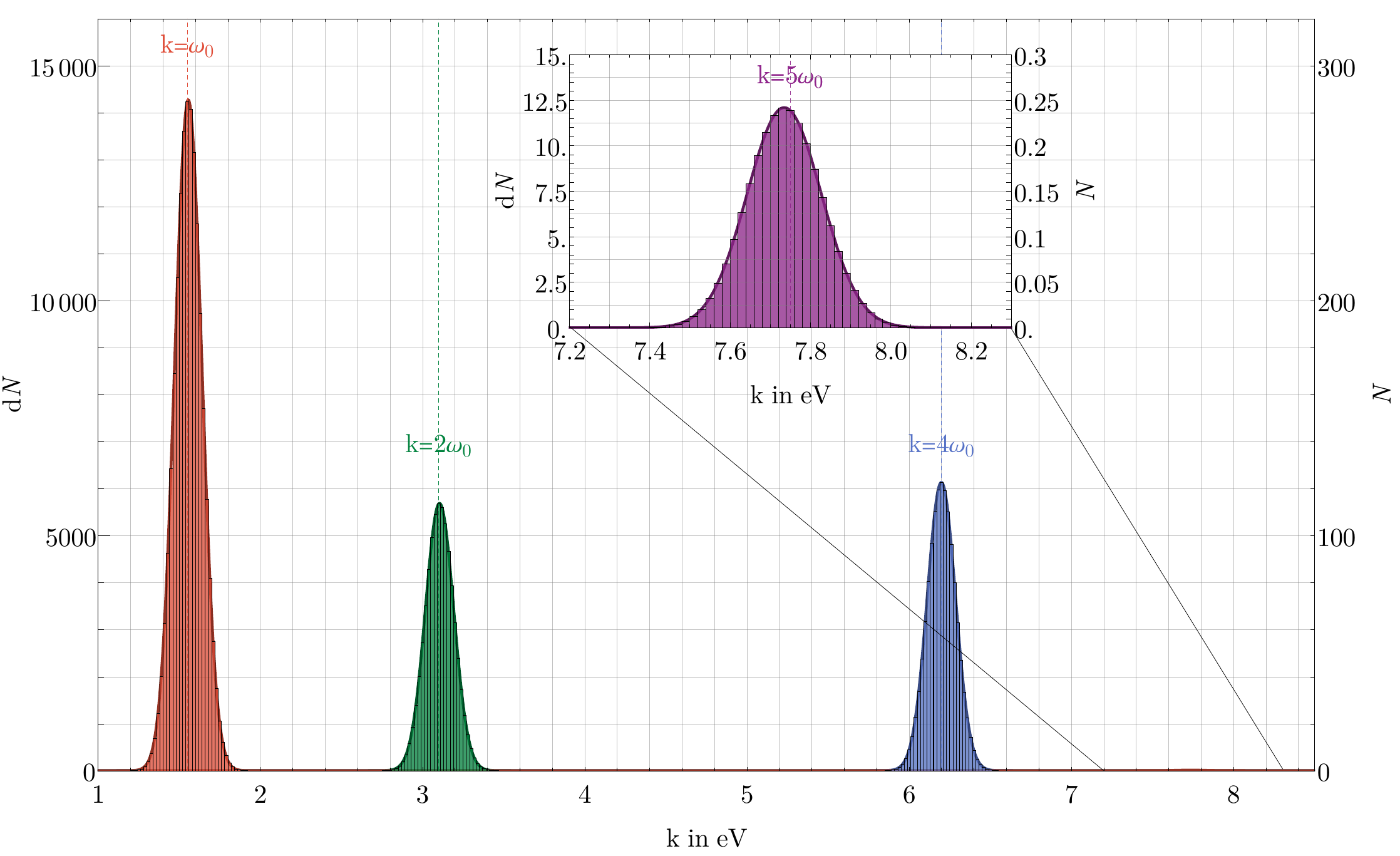}
	\caption{Histogram of $N(4\pi|{\rm k},{\rm k}+\Delta{\rm k})$ in the frequency regime $1\,{\rm eV}\lesssim{\rm k}\lesssim8.3\,{\rm eV}$; the bin range is $\Delta {\rm k}=0.02\,{\rm eV}$ and the signal photon number per bin is given on the right axis. The left axis gives the differential number of signal photons ${\rm d}N$ determined by performing Gaussian fits to the histogram data (solid lines). The integrals of these curves reproduce the signal photon numbers counted in the histograms reasonably well. In the spectral regime highlighted here, the signal photons are predominantly induced at frequencies ${\rm k}\simeq n\omega_0$ with $n\in\{1,2,4\}$ (dashed vertical lines) matching those of the driving laser beams. In addition, we encounter a signal peaked around $5\omega_0$.} 
	\label{fig:Energydistributions}
\end{figure*}

First, we aim at resolving the frequency spectrum of the full signal in detail. The full signal is obtained upon integration of \Eqref{eq:NpA} over all emission directions, such that $\mathcal{A}=4\pi$, and summing over both transverse polarizations $p$.
To this end, we sample the signal photon number $N(4\pi|{\rm k},{\rm k}+\Delta{\rm k})$ with a bin range of $\Delta {\rm k}=0.02\,\text{eV}$.
The results of this analysis are presented as histograms in \Figref{fig:Energydistributions}.
The signal spectrum exhibits four pronounced maxima; the positions of three maxima match the oscillation frequencies of the driving laser fields.  
The additional maximum is centered around ${\rm k}\simeq5\omega_0$.
Adding the contributions of all bins we obtain a total number of $N_{\text{tot}}\simeq5600$ 
signal photons. 

In the present scenario, the positions of all the peaks can be understood in terms of elastic \cite{Heinzl:2006xc,DiPiazza:2006pr,Tommasini:2010fb,Monden:2011,King:2012aw,Dinu:2014tsa,Karbstein:2015xra,Kadlecova:2019dxv,Jeong:2020vpy} and manifestly inelastic \cite{Mckenna:1963,Varfolomeev:1966,Rozanov:1996,Moulin:2002ya,Lundstrom:2005za,DiPiazza:2005jc,Lundin:2006wu,Fedotov:2006ii,Bernard:2010dx,Gies:2014jia,Fillion-Gourdeau:2014uua,Bohl:2015uba,Gies:2017ezf,Aboushelbaya:2019ncg} sum or difference frequency generation processes involving only the oscillation frequencies of the driving laser fields; cf. in particular also \cite{Gies:2017ygp,King:2018wtn}.
The reason is that the pulse duration $\tau$ is much larger than the cycle durations $1/(\nu_i\omega_0)$. It is straightforward to verify that signal photon emission in the formal limit of $\tau\rightarrow \infty$ is indeed restricted to sharp delta peaks at the frequencies of the driving beams and combinations thereof; in this limit the Fourier integral~\eqref{eq:FourierInt} over time results in Dirac delta functions \cite{Gies:2017ygp}.
The finite width of the peaks in frequency space is a consequence of the finiteness of $\tau$, and implies that the frequency selection rules associated with the limit of $\tau\to\infty$ are fulfilled only approximately.

While this is obvious for the quasi-elastic signal photon channels at $\omega_i$,
the signal with frequency ${\rm k}\simeq5\omega_0$ outside the frequency spectra of the driving beams highlighted in the inlay of \Figref{fig:Energydistributions} can be attributed to a sum and difference frequency generation process.

Besides the number of signal photons per bin, \Figref{fig:Energydistributions} shows the differential number of signal photons ${\rm d}N/{\rm dk}$ extracted from these histograms:
upon dividing the signal photon numbers in a given bin by the bin range $\Delta {\rm k}$, we assume their distribution in a given frequency range to be well described by a Gaussian function.
In all cases, the fitted peak values of the Gaussians are close the frequencies $\nu_i\omega_0$.
We find the peak values at $(1.556\pm 4.8 \times 10^{-5}){\rm eV}$, $(3.11 \pm 2.8 \times 10^{-5}){\rm eV}$, $(6.20 \pm 4.9 \times 10^{-6}){\rm eV}$, and $(7.74\pm 4.0 \times 10^{-5}){\rm eV}$. 
The Gaussian standard deviations $\sigma_G$ are $(84.47\pm 0.19)\times10^{-3}\,\text{eV}$ for the $\omega_0$ signal, $(87.01\pm 0.11)\times10^{-3}\,\text{eV}$ for $2\omega_0$, $(86.97\pm 0.02)\times10^{-3}\,\text{eV}$ for $4\omega_0$, and $(92.62\pm 0.18)\times10^{-3}\,\text{eV}$ for $5\omega_0$.
In the interval $\omega_i\pm3\sigma_G$ $99\%$ of the Gaussian distributed signal is located. The values for $\sigma_G$ extracted here are about an order of magnitude smaller than the width of $1.02\,{\rm eV}$ employed to cover the full signal in the previous section. Correspondingly, the above choice should indeed reliably cover the full signal, while being still small enough to prevent an overlap of the signals associated with other frequencies. Besides, the bin size $\Delta{\rm k}$ should be sufficiently small to resolve potential deviations from Gaussian distributions in the spectral domain.

\begin{figure*}[t]
	\centering
	\includegraphics[scale=1]{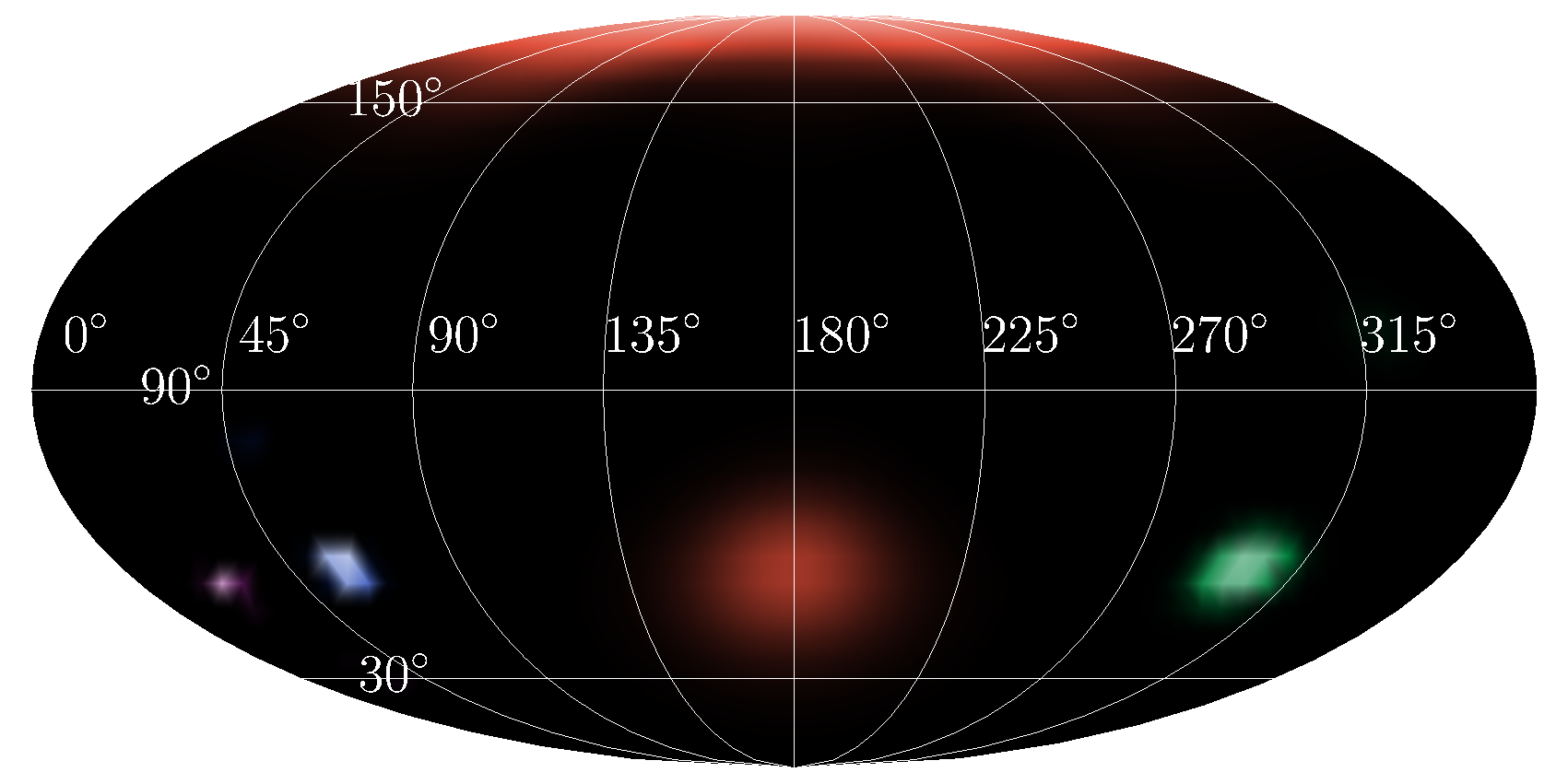}
	\caption{Mollweide plot (longitude $\varphi$, latitude $\vartheta$) of the signal photon density $\rho(\varphi,\vartheta|{\rm k}_{\rm min},{\rm k}_{\rm max})$ highlighting four distinct frequency regimes ${\rm k}_{\rm min}\leq{\rm k}\leq{\rm k}_{\rm max}$ in different colors, namely $1.04\,\mathrm{eV}\leq{\rm k}\leq2.06\,\mathrm{eV}$ (red), $2.59\,\mathrm{eV}\leq{\rm k}\leq3.61\,\mathrm{eV}$ (green), $5.69\,\mathrm{eV}\leq{\rm k}\leq6.71\,\mathrm{eV}$ (blue), and $7.24\,{\rm eV}\leq{\rm k}\leq8.26\,{\rm eV}$ (violet). All four color scales are linear and are normalized to the maximum value in the respective frequency interval.}
	\label{fig:MollSig}
\end{figure*}

The directional distribution of the signal photons of energies ${\rm k}_{\rm min}\leq{\rm k}\leq{\rm k}_{\rm max}$ attainable in a polarization insensitive measurement is encoded in the number density $\rho(\varphi,\vartheta|{\rm k_{min}},{\rm k_{max}})$.
\Figref{fig:MollSig} shows the Mollweide projection of the signal photon density as a function of the azimuthal and polar angles $\varphi$ and $\vartheta$, accounting for four different frequency regimes.
This projection maps the surface of a sphere onto a flat two dimensional chart conserving areas. Obviously, it is not conformal.

The four different frequency regimes are highlighted in different colors. We adopt linear color scales which are normalized to the maximum value in a given frequency regime. 
The brightest areas of a given color mark the dominant emission directions of the signal photons.
As to be expected, the signal photons of frequencies close to $\omega_0$, $3\omega_0$ and $4\omega_0$, respectively, are predominantly emitted in the forward cones of the driving laser pulses featuring the same frequencies: apart from a frequency-$\omega_0$ (red) peak at $\vartheta=180^{\circ}$, we observe three distinct maxima at $\vartheta\approx 54.74^{\circ}$ which are separated by $\approx120^\circ$. These agree with the forward directions of the additional -- from left to right -- $4\omega_0$ (blue), $\omega_0$ (red) and $2\omega_0$ (green) beams.
Additionally, we encounter a $5\omega_0$ (violet) signal at $\varphi\approx 23.26^{\circ}$ and $\vartheta\approx50.85^{\circ}$.

Upon plugging these signal densities into \Eqref{eq:NpA} and integrating over the full solid angle, we obtain the signal photon numbers in the respective frequency regimes. This yields $N(4\pi|1.04\,{\rm eV},2.06\,{\rm eV})\simeq3018.9$ signal photons in the regime around $\omega_0$, $N(4\pi|2.59\,{\rm eV},3.61\,{\rm eV})\simeq1240.8$ around $2\omega_0$, $N(4\pi|5.69\,{\rm eV},6.71\,{\rm eV})\simeq1337.6$ around $4\omega_0$ and $N(4\pi|7.24\,{\rm eV},8.26\,{\rm eV})\simeq2.81$ in the vicinity of $5\omega_0$; cf. also Fig.~\ref{fig:NlA}.

We conclude this section by emphasizing that we have mainly focused on the total numbers of signal photons induced in specific frequency intervals so far and did not address the question of their measurability. This is particularly unclear for the signals at $\omega_0$, $2\omega_0$ and $4\omega_0$ which have been shown to be predominantly emitted into the forward directions of the associated driving beams. 
In the next section we will address this question and assess carefully which signal photon contributions could be isolated or {\it distinguished} from the large background of the driving laser photons in experiment.

\subsection{Discernible signal photons} \label{sec:Signal_and_BG}

To assess if a specific signal can be discerned from the background of the driving laser photons or not, we first have a look at the angular distribution of the latter. Here we have modeled the driving lasers as Gaussian beams.
The far-field angular decay of a Gaussian beam made up of ${\cal N}_i$ photons can be expressed as \cite{Karbstein:2019oej}
\begin{equation}
 {\rm d}^2 \mathcal{N}_i (\varphi,\vartheta) = {\rm d}^2\Omega\,\frac{{\cal N}_i}{2\pi}(\omega_i w_i)^2\,{\rm e}^{-(\omega_iw_i)^2\Theta_i^2(\varphi,\vartheta)/2}\,,
 \label{eq:d2calNi}
\end{equation}
where the angle $\Theta_i(\varphi,\vartheta)$ parameterizes the angular decay of the laser photons measured from the forward beam axis $\hat{\pmb\kappa}_i$. As we consider Gaussian beams which feature a rotational symmetry around the beam axis, a single angle parameter is sufficient.
In the present scenario, we have $\Theta_0=\vartheta - \pi$ and $\Theta_i=-{\rm arccos}\{\cos\vartheta[\cos\alpha+\cos(\varphi-2\pi\frac{4-i}{3})\sin\alpha]\}$ for beams $i\in\{1,2,3\}$.
Moreover, recall that we have $\mathcal{N}_i=WA_i^2/(\nu_i\omega_0)$ and $\omega_i w_i=2\pi\nu_i$; cf. Sec.~\ref{sec:amplitudes_losses}.
In the following, we use the notation $ {\rm d}^2\mathcal{N}(\varphi,\vartheta)=\sum_{i=0}^3 {\rm d}^2\mathcal{N}_i(\varphi,\vartheta)$ for the differential number of photons ${\cal N}=\sum_{i=1}^3{\cal N}_i$ constituting all laser beams.

Since we have assumed that all laser beams are focused to the same waist $w_i=\lambda$, the far-field angular divergences of the beams scale as $\sim1/\nu_i$.
This implies that the beam with the largest value of $\nu_i$ features the smallest far-field divergence. At the same time, the effective extent of the interaction region, determining the far-field divergences of the quasi-elastically scattered signal photons should be similar for all individual beams.
The steeper decay of the laser photons constituting the $4\omega_0$ background suggests that the signal photons arising from the $4\omega_0$ beam should be more easily detectable than the analogous contributions for the other beams.

To illustrate the major challenge of signal-to-background separation we emphasize the huge background provided by the photons constituting the driving laser pulses.
Their total number per shot is as large as ${\cal N}\simeq5.3\times10^{20}$ to be contrasted with the number of ${\cal O}(10^3)$ signal photons achievable in this setup; cf. Sec.~\ref{sec:amplitudes_losses} above.

\begin{figure*}[t]
	\centering
	\includegraphics[scale=1]{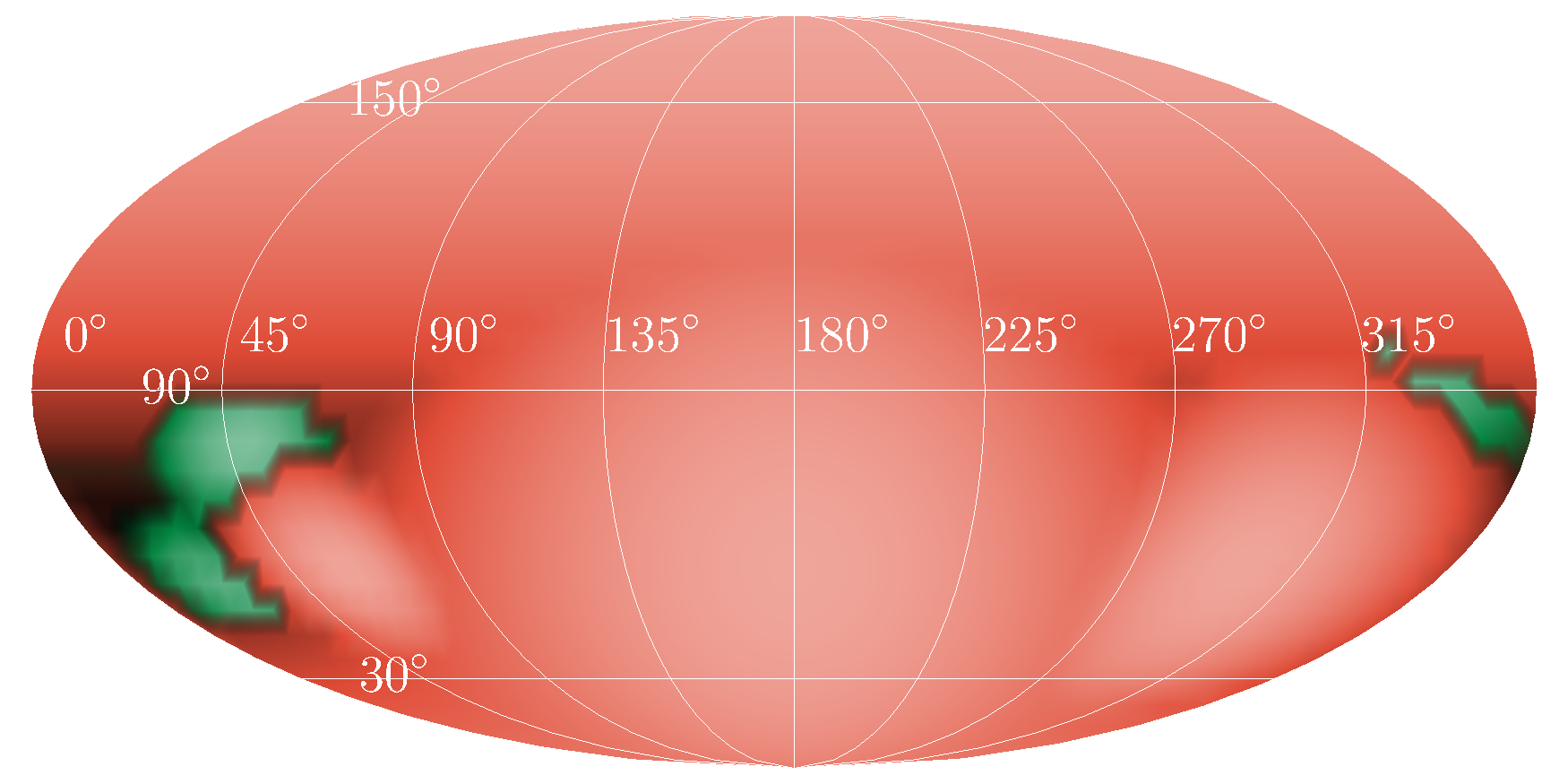}
	\caption{Directional characteristics (longitude $\varphi$, latitude $\vartheta$) of the differential number of driving laser photons ${\rm d}^2\mathcal{N}(\varphi,\vartheta)$ constituting the background and the associated signal photon density $\rho(\varphi,\vartheta|0,\infty)$. Angular regions where the signal dominates over the background are colored in green; regions where the background dominates in red.
	In each color regime we adopt a logarithmic color scale normalized to the maximum value.}
	\label{fig:MollDiff}
\end{figure*}

Figure~\ref{fig:MollDiff} depicts on the directional characteristics of both the signal and driving laser photons on a logarithmic scale.
While the background of the laser photons indeed dominates in most directions, the signal surpasses the background in certain angular regions.
This already hints at the principle possibility of measuring the signal over the background in certain directions -- though it is unclear at this point whether there are enough discernible signal photons yielding a sufficient statistics in a concrete experiment.

The answer of this question requires a more detailed study.
For this, we proceed to a frequency-resolved analysis.
More specifically, we search for discernible signal photons in the four distinct frequency regimes around $\omega_0$, $2\omega_0$, $4\omega_0$ and $5\omega_0$ introduced above.

\subsubsection*{$\omega_0$ regime}

Isolating the $\omega_0$ signal from the background seems particularly challenging: the signal photon density $\rho(\varphi,\vartheta|{\rm k_{min}},{\rm k_{max}})$ features just two peaks in the frequency regime delimited by ${\rm k_{min}=1.02\,eV}$ and ${\rm k_{max}=1.05\,eV}$, both of which are coinciding with the forward directions of the two driving laser beams of frequency $\omega_0$.
At the same time, exactly these beams come with the largest far-field divergences. Besides, particularly due to the large energy put into beam $0$, the number of background photons is maximal in this frequency regime.
Though it might be an option to discern at least parts of the signal from the background by advanced detection techniques, based on analyses of the decay behavior or polarization details of both the background and the signal photons, here we proceed to the other frequency regimes suggesting more easily accessible signals; cf. below.

\subsubsection*{$2\omega_0$ regime}

\begin{figure*}[t]
	\centering
	\includegraphics[scale=1]{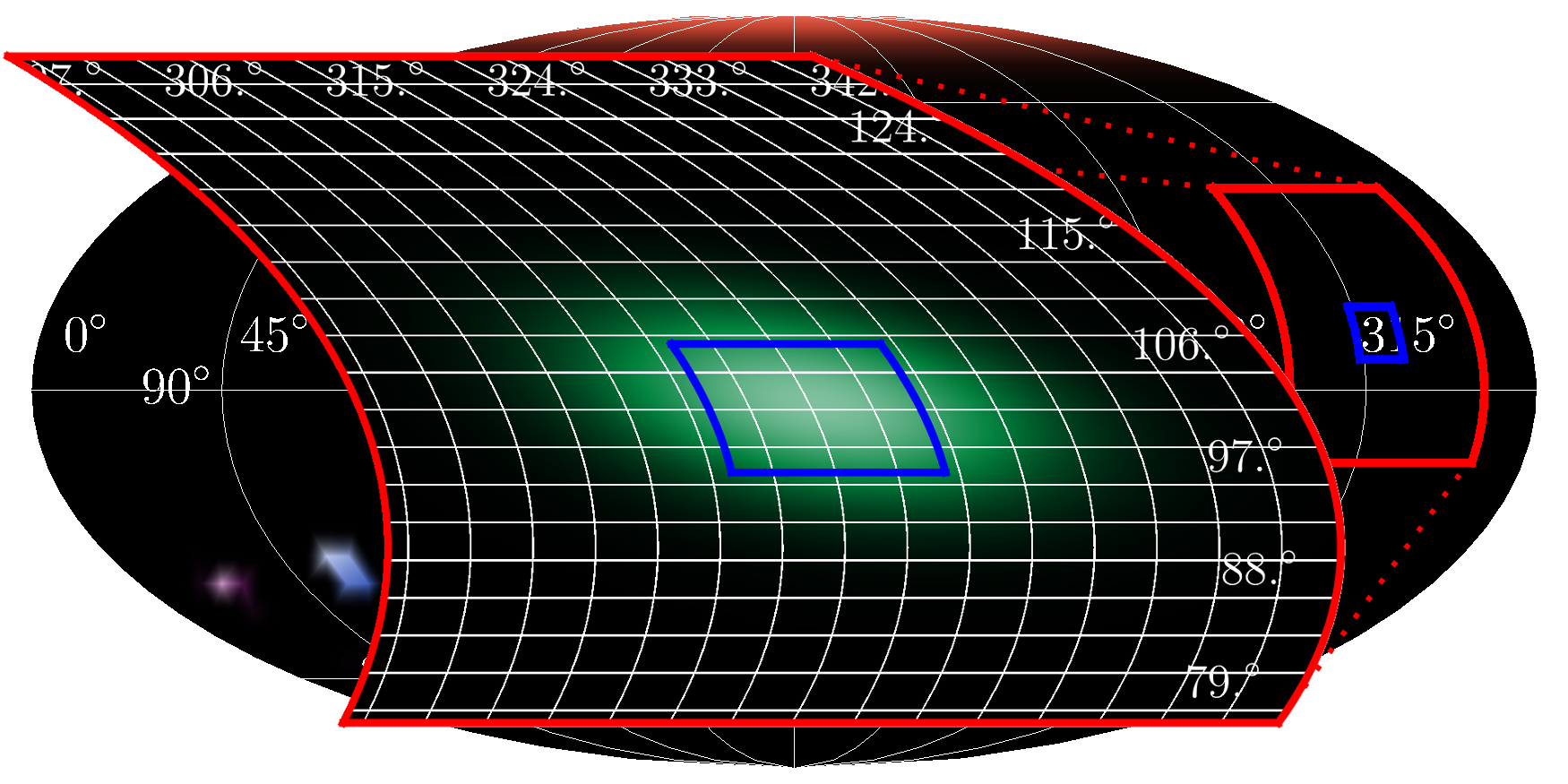}
	\caption{Mollweide plot (longitude $\varphi$, latitude $\vartheta$) of the signal photon density $\rho(\varphi,\vartheta|{\rm k}_{\rm min},{\rm k}_{\rm max})$.
	We highlight the relevant angular domain (marked by a red frame) for signal photon energies in the regime constrained by $2.59\,\mathrm{eV}\leq{\rm k}\leq3.61\,\mathrm{eV}$; the linear green color scale is normalized to its maximum in its frame.
	The blue frame marks the angular region for which the number of discernible signal photons quoted in the main text is determined.}
	\label{fig:MollZoomOm2}
\end{figure*}

Next, we focus on the frequency regime centered around $2\omega_0\simeq3.1\,{\rm eV}$ and constrained by ${\rm k_{min}}=2.59\,\text{eV}$ and ${\rm k_{max}}=3.61\,\text{eV}$.
As is clearly visible in Fig.~\ref{fig:MollZoomOm2}, using the numerical Newton method we identify a local maximum of the signal photon density $\rho(\varphi,\vartheta|{\rm k_{min}},{\rm k_{max}})$ at $(\varphi,\vartheta)\simeq(317.65^{\circ},101.77^{\circ})$.
By comparison with Fig.~\ref{fig:MollSig}, this particular maximum is clearly separated from the forward beam axis of the driving $2\omega_0$ beam constituting the main background in this specific frequency regime.
For an estimate of the quantitative number of signal photons, we limit ourselves to the angular region $\mathcal{A}^{(2\omega_0)}=\{ (\varphi,\vartheta)|\varphi\in[314^{\circ},324^{\circ}],\vartheta\in[96^{\circ},106^{\circ}]\}$ marked by the blue frame in Fig.~\ref{fig:MollZoomOm2}.
An integration over this angular region results in $N(\mathcal{A}^{(2\omega_0)}|2.59\,\text{eV},3.61\,\text{eV})\simeq62$ 
signal photons per shot.
For completeness, we note that this value essentially constitutes the full number $N(\mathcal{A}^{(2\omega_0)}|0,\infty)$ of signal photons emitted into this angular regime. 

Upon numerically integrating \Eqref{eq:d2calNi} for the $2\omega_0$ beam $i=2$ over the same angular interval, we find $\mathcal{N}(\mathcal{A}^{(2\omega_0)}) \simeq 0.01$ 
background photons per shot. Of course, the other driving beams do not induce a background in this frequency regime.
This analysis implies that essentially all $\simeq62$ photons with the considered directional characteristics are signal photons, which can thus be clearly distinguished from the background.

\subsubsection*{$4\omega_0$ regime}

\begin{figure*}[t]
	\centering
	\includegraphics[scale=1]{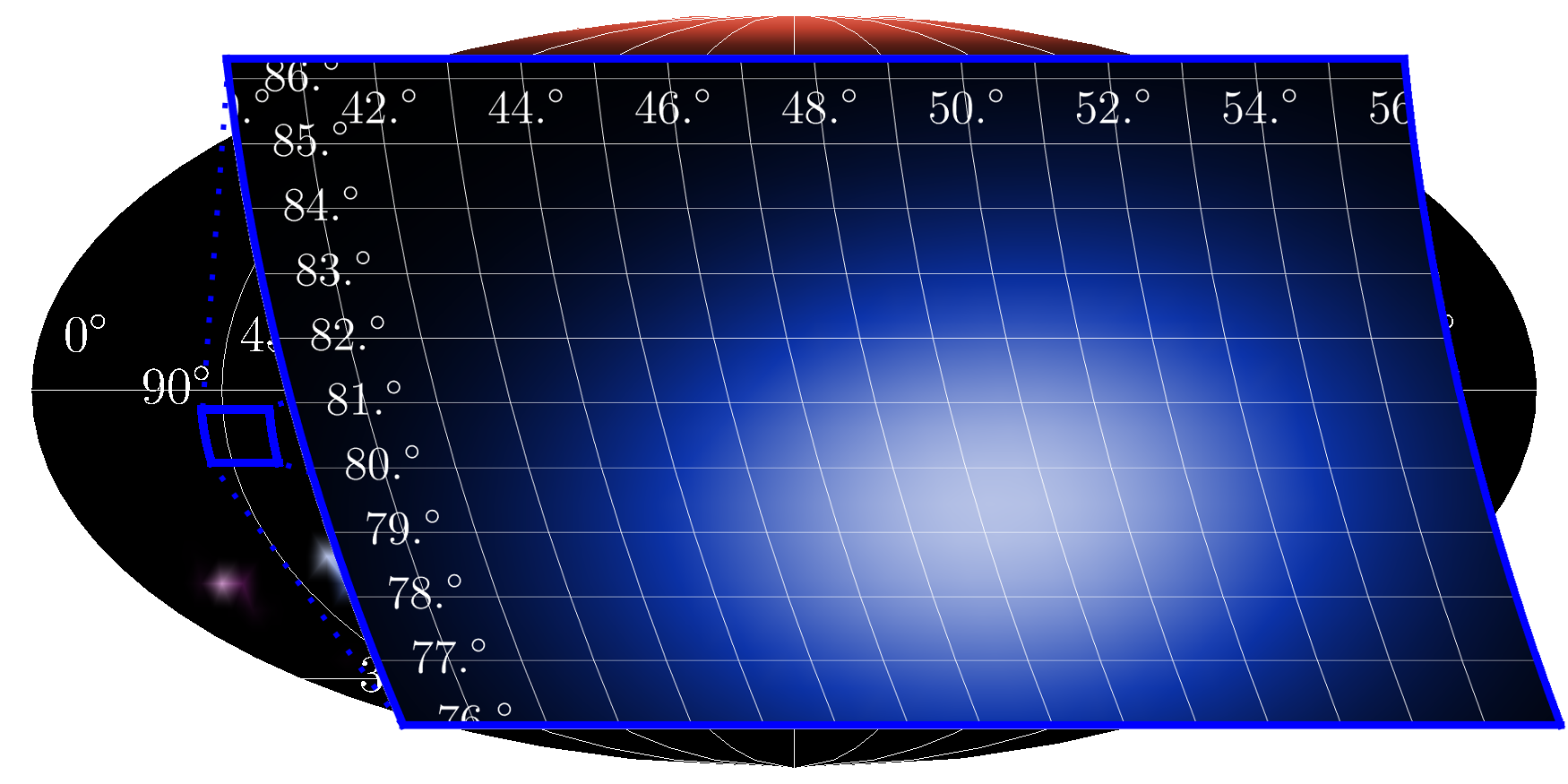}
	\caption{Mollweide plot (longitude $\varphi$, latitude $\vartheta$) of the signal photon density $\rho(\varphi,\vartheta|{\rm k}_{\rm min},{\rm k}_{\rm max})$.
	We highlight the relevant angular domain (marked by a blue frame) for signal photon energies in the regime constrained by $5.69\,\mathrm{eV}\leq{\rm k}\leq6.71\,\mathrm{eV}$; the linear blue color scale is normalized to its maximum in its frame.
	The blue frame marks the angular region for which the number of discernible signal photons quoted in the main text is determined.}
	\label{fig:MollZoomOm3}
\end{figure*}

Further, we turn to  $4\omega_0$ frequency regime constrained by ${\rm k_{min}}=5.69\,\text{eV}$ and ${\rm k_{max}}=6.71\,\text{eV}$.
Also in this regime we search for a local maximum of the signal photon density besides the dominant one in the forward cone of driving  $4\omega_0$ laser beam $i=3$.
A numerical analysis of the signal photon density $\rho(\varphi,\vartheta|{\rm k_{min}},{\rm k_{max}})$ utilizing the Newton method allows us to identify a local maximum with the desired properties at $(\varphi,\vartheta)\simeq(49.43^{\circ},79.44^{\circ})$.
See Fig.~\ref{fig:MollZoomOm3} for a graphical illustration of the signal photon density in the relevant angular regime.

An integration of the signal density over the area $\mathcal{A}^{(4\omega_0)}=\{ (\varphi,\vartheta)|\varphi\in[40^{\circ},56^{\circ}],\vartheta\in[76^{\circ},86^{\circ}]\}$ highlighted in Fig.~\ref{fig:MollZoomOm3} results in $N(\mathcal{A}^{(4\omega_0)}) \approx 129$ 
signal photons per shot. We have explicitly checked that this is the full number of signal photons emitted into this angular regime; there are no signal photons of other frequencies.
For comparison, in the same angular regime we find $\mathcal{N}(\mathcal{A}^{(4\omega_0)}) \simeq 0.006$
driving laser photons of frequency $4\omega_0$ per shot constituting the background.

\subsubsection*{$5\omega_0$ regime}

As noted in Sec.~\ref{sec:distribution}, apart from the signals just discussed, we also identify a signal outside the frequencies of the driving laser beams in the energy regime of $7.24\,{\rm eV}\leq{\rm k}\leq8.26\,{\rm eV}$, featuring a peak at about $5\omega_0\simeq7.75\,{\rm eV}$.
A numerical analysis of the signal photon density $\rho(\varphi,\vartheta|{\rm k}_{\rm min},{\rm k}_{\rm max})$ in this frequency regime unveils the existence of two pronounced maxima signalizing two different main signal photon emission directions at (A): $(\varphi,\vartheta)\simeq(23.26^{\circ},50.85^{\circ})$
and (B): $(\varphi,\vartheta)\simeq(31.32^{\circ},31.01^{\circ})$, respectively.

\begin{figure*}[t]
	\centering
	\includegraphics[scale=1]{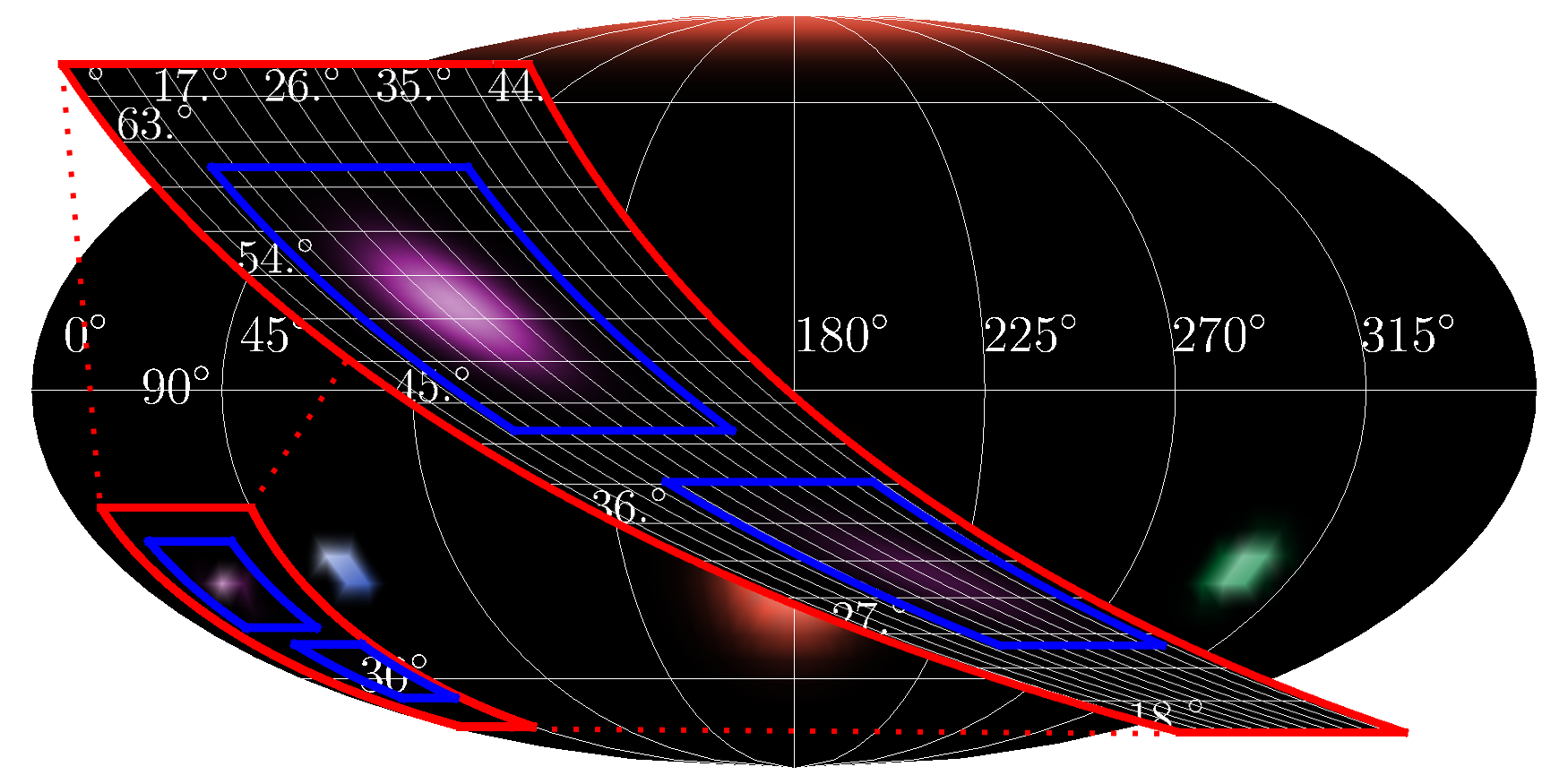}
	\caption{Mollweide plot (longitude $\varphi$, latitude $\vartheta$) of the signal photon density $\rho(\varphi,\vartheta|{\rm k}_{\rm min},{\rm k}_{\rm max})$.
	We highlight the relevant angular domain (marked by a red frame) for signal photon energies in the regime constrained by $7.24\,\mathrm{eV}\leq{\rm k}\leq8.26\,\mathrm{eV}$; the linear violet color scale is normalized to its maximum in its frame.
	The two blue frames mark the angular regions for which the number of discernible signal photons quoted in the main text is determined.}
\label{fig:MollZoomOm4}
\end{figure*}

In \Figref{fig:MollZoomOm4} we illustrate the signal photon density in the relevant angular areas.
The two maxima are located in the two angular areas marked by blue frames.
Maximum (A) is located in the upper blue frame delimiting the area $\mathcal{A}^{(5\omega_0,A)}=\{(\varphi,\vartheta)|\varphi\in[14^{\circ},35^{\circ}],\vartheta\in [42^{\circ},60^{\circ}]\}$, and maximum (B) in the lower frame the area $\mathcal{A}^{(5\omega_0,B)}=\{(\varphi,\vartheta)|\varphi\in[21^{\circ},42^{\circ}],\vartheta\in [25^{\circ},38^{\circ}]\}$.

Upon integration of the signal photon density over the two angular regimes $\mathcal{A}^{(5\omega_0,A)}$ and $\mathcal{A}^{(5\omega_0,B)}$, we obtain $N(\mathcal{A}^{(5\omega_0,A)})\approx 2.3$
and $N(\mathcal{A}^{(5\omega_0,B)})\approx 0.5$ 
signal photons per shot. These values essentially agree with the frequency-unresolved numbers of signal photons emitted into the same angular regions, signaling the presence of $5\omega_0$ photons only.
We emphasize once again that there is no genuine laser photon background in the $5\omega_0$ frequency regime since the driving laser fields only contain frequencies in the vicinity of $\omega_0$, $2\omega_0$ and $4\omega_0$.
By contrast, we do not find any significant signal at higher harmonic frequencies.

We summarize the quantitative findings from the preceding sections in Tab.~\ref{tab:photon_numbers}.
This table features the prospective signal photon numbers and numbers of driving laser photons constituting the background for various frequencies and emission directions. 

\begin{table*}[t]
 \centering
 \caption{Prospective numbers of signal photons $N$ and driving laser photons $\mathcal{N}$ in different energy regimes and angular emission areas ${\cal A}$. See the main text for the definitions of the areas as well as further details.}
 \label{tab:photon_numbers}
	\begin{tabular}{p{0.22\textwidth}<{\centering}p{0.22\textwidth}<{\centering}p{0.22\textwidth}<{\centering}p{0.22\textwidth}<{\centering}}
	\hline \hline 
	  & $2.59{\rm\, eV}\leq{\rm k}\leq3.61{\rm\, eV}$ & $5.69\,\text{eV}\leq{\rm k}\leq 6.71\,\text{eV}$ & $7.24\,\text{eV}\leq{\rm k}\leq 8.26\,\text{eV}$ \\ 
	  & cf. \Figref{fig:MollZoomOm2} & cf. \Figref{fig:MollZoomOm3} & cf. \Figref{fig:MollZoomOm4} \\ \hline
	  $N(4\pi)$ & 1240.80 & 1337.67 & 2.81 \\
	  $\mathcal{N}(4\pi)$ & $6.98\times10^{19}$ & $1.75\times10^{19}$ & 0.00 \\
	  $N(\mathcal{A}^{(2\omega_0)})$ & 62.02 & 0.00 & 0.00 \\
	  $\mathcal{N}(\mathcal{A}^{(2\omega_0)})$ & $10.13\times10^{-3}$ & 0.00 & 0.00 \\
	  $N(\mathcal{A}^{(4\omega_0)})$ & 0.00 & 129.40 & 0.00 \\
	  $\mathcal{N}(\mathcal{A}^{(4\omega_0)})$ & 0.00 & $5.91\times10^{-3}$ & 0.00 \\
	  $N(\mathcal{A}^{(5\omega_0,A)})$ & 0.00 & $8.19\times 10^{- 5}$ & 2.31 \\
	  $N(\mathcal{A}^{(5\omega_0,B)})$ & 0.00 & 0.00 & 0.46 \\
	  \hline \hline
	\end{tabular}

\end{table*}

\section{Channel analysis}\label{sec:channelanalysis}

In Sec.~\ref{sec:scen} we have identified several promising signals and demonstrated that they are, in principle, discernible against the background of the photons of the driving laser beams.
To obtain these results we have relied on a rather time consuming and brute force numerical evaluation of the signal photon density $\rho_{(p)}(\varphi,\vartheta|{\rm k}_{\rm min},{\rm k}_{\rm max})$.
This quantity encodes information about all possible single photon emission processes mediated by quantum vacuum fluctuations in the macroscopic field driving the effect.
In order to resolve different frequency regimes within this approach, we have evaluated $\rho_{(p)}(\varphi,\vartheta|{\rm k}_{\rm min},{\rm k}_{\rm max})$ for various values of ${\rm k}_{\rm min}$ and ${\rm k}_{\rm max}$.

Subsequently, we demonstrate how these results can be obtained with considerably less computational efforts, using the findings of the previous section~\ref{sec:scen} as benchmarks for the new analysis carried out here.
A channel analysis for various three-pulse setups has been performed in \cite{Gies:2017ezf,King:2018wtn}.

\subsection{Tracing the microscopic origin of the signal}\label{sec:tracing}

Our starting point is the expression for the signal photon density in \Eqref{eq:rho_result}, with the modulus squared of the zero-to-single signal transition amplitude given by \Eqref{eq:SpSpstar}.
Interchanging the integration over energy and the summation over $\ell$ and $\ell'$, the signal photon density can be expressed as
\begin{equation}
\rho_{(p)}(\varphi,\vartheta|{\rm k}_{\rm min},{\rm k}_{\rm max}) = \sum_{\ell,\ell'} \rho_{(p);\ell,\ell'}(\varphi,\vartheta|{\rm k}_{\rm min},{\rm k}_{\rm max})\,,
\label{eq:rhosum}
\end{equation}
where the sum runs over all sets $\ell=\{i,j,l\}$, $\ell'=\{i',j',l'\}$ and we have defined
\begin{align}
 &\rho_{(p);\ell,\ell'}(\varphi,\vartheta|{\rm k}_{\rm min},{\rm k}_{\rm max}) \nonumber\\
 &\quad\quad= \frac{1}{(2\pi)^3} \int_{\mathrm{k}_{\rm min}}^{\mathrm{k}_{\rm max}} \mathrm{dk}\,{\rm k}^2\, {\rm Re}\bigl\{S_{(p);\ell}(\mathbf{k})S^*_{(p);\ell'}(\mathbf{k})\bigr\} .
 \label{eq:rholl'}
\end{align}
Accordingly, we introduce the signal photon number $N_{(p);\ell,\ell'}({\cal A}|{\rm k}_{\rm min},{\rm k}_{\rm max})$ following from \Eqref{eq:rholl'} upon integration over the solid angle interval $\cal A$.
Obviously, $N_{(p)}=\sum_{\ell,\ell'}N_{(p);\ell,\ell'}$; cf. \Eqref{eq:NpA}.

While only the sum over all sets $\ell$ and $\ell'$ constitutes the physical density -- off-diagonal terms with $\ell\neq\ell'$ may even be negative -- this representation provides us with a convenient means to assess the importance of individual contributions constituting the full density.
As demonstrated below, it allows us to straightforwardly identify the single interaction processes inducing a given vacuum emission channel.   
This information can then be utilized to optimize the signal photon yield in this channel, e.g., by changing the partitioning of the total available laser energy into the different driving beams. 

A closer look at the structure of $S_{(p);ijl}$ in Eqs.~\eqref{eq:S_p}-\eqref{eq:geo_func} unveils certain symmetries which can be employed to reduce the number of contributions to be evaluated explicitly.
Most obviously, the Fourier integral~\eqref{eq:FourierInt} is completely symmetric under permutations of the indices $i$, $j$, $l$.
At the same time, the function~\eqref{eq:geo_func} which encodes the directional characteristics of the interacting fields only exhibits a reduced symmetry: it is symmetric under exchange of the first two indices, i.e., $g_{(p);ijl}(\hat{\bf k}\bigr)=g_{(p);jil}(\hat{\bf k}\bigr)$, and vanishes if these two indices agree, $g_{(p);iil}(\hat{\bf k}\bigr)=0$.
Correspondingly, we have $S_{(p);ijl}=S_{(p);jil}$ and $S_{(p);iil}=0$.
This implies that the only non-vanishing contributions with exactly two identical indices can be expressed in terms of $S_{(p);iji}$, where $i\neq j$.
On the other hand, given that all indices are different, i.e., $i\neq j\neq l$, the only independent contributions arise from $S_{(p);ijl}$.

Recall that when studying a collision scenario involving $n+1$ driving laser fields each of the indices runs from $0$ to $n$.
In this case the above considerations result in $n(n+1)$ independent non-vanishing contributions to $\sum_\ell S_{(p);\ell}$ due to terms for which all three indices disagree, and another $n(n+1)$ ones from terms with two identical indices. 
Terms with three indices identical vanish completely.
Of course, the very same considerations apply to $\sum_{\ell'} S^*_{(p);\ell'}$.

In the next step, we focus on an individual contribution with fixed $\ell=\{i,j,l\}$, in order to identify the signal photon frequency associated with this channel.
To this end it is helpful to consider the formal limit of $\tau\rightarrow \infty$ corresponding to the collision of monochromatic laser beams with constant temporal envelope; cf. also \cite{Gies:2017ygp}.
As noted in Sec.~\ref{sec:distribution}, in this limit the temporal integration in $\mathcal{I}_{ijl}$  results in a delta function ensuring the signal photon energy $\rm k$ to be fully determined by the oscillation frequencies $\omega_i=\nu_i\omega_0$ of the driving laser beams,
\begin{equation}
{\rm k} \to | \pm \omega_i \pm \omega_j \pm \omega_l |\,, \label{eq:omega_S}
\end{equation}
where each sign can occur separately.
A positive (negative) sign corresponds to the emission (absorption) of a photon from the respective driving laser beam.
For instance, the contribution with $\ell=\{0,3,0\}$ triggers signal photon frequencies of $2\omega_0$, $4\omega_0$ and $6\omega_0$.
We emphasize that, in order to obtain a sizable signal photon contribution of \Eqref{eq:rholl'} at a given frequency, both factors ${\cal S}_{(p);\ell}$ and ${\cal S}^*_{(p);\ell'}$ in \Eqref{eq:rholl'} need to support this frequency.
For finite pulse durations $\tau$, the selection rules in \Eqref{eq:omega_S} hold only approximately. 
Nevertheless, this approximation is accurate as long as the propagation directions of the driving beams are sufficiently different and $\tau\omega_0\gg1$; cf. Sec.~\ref{sec:distribution}. This is rather generic for scenarios envisioning the collision of several quasi-monochromatic high-intensity laser pulses. In this case the finite pulse envelopes blur the selection rules~\eqref{eq:omega_S}, and the sharp delta peaks resulting from the temporal integration in $\mathcal{I}_{ijl}$ for $\tau\to\infty$ are replaced by Gaussian peaks of a finite spectral width scaling as $\sim1/\tau$.

In the scenario considered here we have $\omega_0\tau\simeq 59\gg1$, or equivalently $1/\tau\simeq\omega_0/59\ll\omega_0$, which clearly hints at the fact that the spectral width of these Gaussian peaks is much smaller than the spectral separation of any two oscillation frequencies $\omega_i$ of the driving laser fields.
This is in line with the findings of Sec.~\ref{sec:distribution}:
here Gaussian fits to the various peaks encountered in the differential number of signal photons ${\rm d}N/{\rm dk}$ resulted in Gaussian standard deviations of the same magnitude.
For the largest extracted frequency width ($\sigma=0.0926\,{\rm eV}$), the associated full peak-width measured at $1\%$ of the peak-maximum is given by $\Delta\omega=4\sqrt{\log 100}\,\sigma\, \simeq0.513\,\omega_0$.

Hence, instead of analyzing the signal photon spectrum by explicitly segmenting the signal photon density in the spectral domain as done in Sec.~\ref{sec:distribution}, we can resolve its frequency spectrum by limiting the sum in \Eqref{eq:rhosum} to the relevant channels which induce non-vanishing contributions in the vicinity of a given frequency.

If in addition the projections of the oscillation-period-averaged intensity profiles on the beam axes of all other beams vary on scales much larger than the wavelengths of the beams, also the wave vector of the signal photon $\bf k$ should -- to a good approximation -- be determined by the wave vectors of the photons comprising the beams in the plane-wave limit $\{\tau,w_i\}\to\infty$ where focusing effects can be neglected. In this case, we have
\begin{equation}
 {\bf k}\simeq {\bf k}_{\rm pw} =\pm\omega_i\hat{\pmb{\kappa}}_i\pm\omega_j\hat{\pmb{\kappa}}_j\pm\omega_k\hat{\pmb{\kappa}}_k\,, \label{eq:k}
\end{equation}
where ${\bf k}_{\rm pw}$ denotes the corresponding  wave vector in the plane-wave limit.
Under these conditions, only contributions with $i\neq j\neq l$, i.e., manifestly inelastic signal photon contributions arising from the mixing of three different driving waves, may result in clearly discernible signals.

To demonstrate this we first focus on the complementary case where two indices agree.
If, e.g., $i=l$ and $i\neq j$, \Eqref{eq:omega_S} predicts signals either at ${\rm k}\simeq\omega_j$ or at ${\rm k}\simeq\omega_j\pm2\omega_i$.
However, at the same time, \Eqref{eq:k} implies ${\bf k}\simeq\pm\omega_j \hat{\pmb{\kappa}_j}$ or ${\bf k}\simeq\pm\omega_j\hat{\pmb{\kappa}}_j\pm2\omega_i\hat{\pmb{\kappa}}_i$.
From these findings it is obvious that only the conditions ${\rm k}=|{\bf k}|=\omega_j$ are compatible with each other for generic values of $\omega_i$ and $\omega_j$ as well as non-collinear $\hat{\pmb{\kappa}}_i$ and $\hat{\pmb{\kappa}}_j$ as considered here.
On the other hand, signal photons fulfilling ${\rm k}=|{\bf k}|=\omega_j$ are expected to be predominantly emitted in the forward direction $\hat{\pmb{\kappa}}_j$ of the driving laser beam of frequency $\omega_j$, rendering their experimental detection very challenging. 
For completeness, note the principle possibility of a quantum reflection signal in the opposite direction, which is completely negligible for the present scenario where focusing effects are found to be subleading \cite{Gies:2013yxa}.

As an illustrative example we determine the signal photon number associated with the channel $\ell=\ell'=\{0,3,0\}$, yielding $N_{030,030}(4\pi|5.69\,{\rm eV}, 6.71\,{\rm eV})\simeq463.46$ photons per shot in the frequency regime around $4\omega_0$.
This signal is peaked at $\hat{\boldsymbol{\kappa}}_3$.
On the other hand, we find $N_{030,030}(4\pi|0,\infty)\sim463.46$, such that -- as expected, and in line with the arguments given above -- obviously no signal photons at other frequencies contribute to this channel.

In the remainder of this section our focus is on the manifestly inelastic signal photon contributions associated with the 12 independent combinations to the signal photon amplitude characterized by $i\neq j\neq l$.
In particular, we aim at verifying the signals arising in the angular regions $\mathcal{A}^{(2\omega_0)}$, $\mathcal{A}^{(4\omega_0)}$ and $\mathcal{A}^{(5\omega_0,A/B)}$ introduced in Sec.~\ref{sec:Signal_and_BG}.
This allows us to explicitly restrict our analysis to channels giving rise to signal photons of the desired energy and wave vectors pointing in the respective directions.  

Tracking the $2\omega_0$ signal in $\mathcal{A}^{(2\omega_0)}$, we analyze all permutations of the indices $0$, $1$ and $2$.
Microscopically, we expect this signal to arise from a process involving the merging of two laser photons from beams $0$ and $2$, respectively, and the absorption of a laser photon of frequency $\omega_1$ from beam $1$.
Resorting to the plane-wave approximation, this results in a signal photon wave vector of modulus $|\mathbf{k}_{\rm pw}| \approx 2.201 \omega_0$ pointing at $(\varphi,\vartheta)=(319.107^{\circ},101.07^{\circ})$. Obviously this value is compatible with 
the condition $|{\rm k}-|{\bf k}_{\rm pw}||<\Delta\omega$ and ${\rm k}\simeq 2\omega_0$. It thus allows for a nonvanishing signal in this parameter regime.
A comparison with the dominant emission direction determined numerically in Sec.~\ref{sec:Signal_and_BG} unveils an excellent agreement; the relative differences in the longitude and latitude are below $1\%$. The signal associated with this channel is found to be peaked around ${\rm k}\simeq 2.015\omega_0$.

Restricting the sums over $\ell$ and $\ell'$ in \Eqref{eq:rhosum} to all possible permutations of the indices $0$, $1$ and $2$ and integrating over the full solid angle we obtain $N(4\pi|0,\infty)\simeq78$
signal photons per shot, while an explicit restriction to the angular region $\mathcal{A}^{(2\omega_0)}$ results in $N(\mathcal{A}^{(2\omega_0)}|0,\infty)\simeq62$.
To study the importance of the individual contributions to this sum, in Tab.~\ref{tab:subchannels} (a) we explicitly list the contributions of all nine terms constituting the signal photon number in this parameter regime.
This also allows us to assess the relative importance of off-diagonal terms with $\ell\neq\ell'$.
Exactly the same number is obtained when the frequency regime is in addition restricted to $2.59\,{\rm eV}\leq{\rm k}\leq3.61{\rm eV}$.
A comparison with the analogous number extracted in Sec.~\ref{sec:Signal_and_BG} establishes that all signal photons scattered into this parameter regime are indeed emerging from the microscopic process $\omega_0-\omega_1+\omega_2\to{\rm k}$.

\begin{table*}[t]
	\caption{Comparison of various contributions $N_{\ell,\ell'}({\cal A}^{(n\omega_0)}|0,\infty)$ to the signal photon number. Here, we highlight several photon emission channels resulting in manifestly inelastically scattered signal photons of frequency ${\rm k}\simeq n\omega_0$ with $n\in\{2,4,5\}$.}
    \parbox{0.95\columnwidth}{ (a): ${\rm k}\simeq2\omega_0$  \hfill (b): ${\rm k}\simeq4\omega_0$ \vspace{5pt} }\\
	\parbox{0.95\columnwidth}{\begin{tabular}{p{0.1\textwidth}<{\centering}p{0.1\textwidth}<{\centering}p{0.1\textwidth}<{\centering}p{0.1\textwidth}<{\centering}}
        \hline \hline
        \diagbox{$\ell$}{$\ell'$} & $0\;1\;2$ & $1\;2\;0$ & $0\;2\;1$  \\ \hline
        $0\;1\;2$ & 1.557 & 0.327 & 6.516  \\
        $1\;2\;0$ & 0.327 & 2.143 & 5.122 \\
        $0\;2\;1$ & 6.516 & 5.122 & 34.394 \\
        \hline \hline
	\end{tabular}
	}\;
   \parbox{0.950\columnwidth}{\begin{tabular}{p{0.1\textwidth}<{\centering}p{0.1\textwidth}<{\centering}p{0.1\textwidth}<{\centering}p{0.1\textwidth}<{\centering}}
        \hline \hline
        \diagbox{$\ell$}{$\ell'$} & $0\;1\;3$ & $1\;3\;0$ & $0\;3\;1$   \\ \hline
        $0\;1\;3$ & 0.437 & 0.754 & 4.917  \\
        $1\;3\;0$ & 0.754 & 12.814 & 18.851 \\
        $0\;3\;1$ & 4.917 & 18.851 & 67.105 \\
        \hline \hline
	\end{tabular}
	}\;\;
	\parbox{0.95\columnwidth}{\vspace{5pt} (c): ${\rm k}\simeq5\omega_0$  \hfill (d): ${\rm k}\simeq5\omega_0$  }\\
   \parbox{0.95\columnwidth}{\begin{tabular}{p{0.1\textwidth}<{\centering}p{0.1\textwidth}<{\centering}p{0.1\textwidth}<{\centering}p{0.1\textwidth}<{\centering}}
        \hline \hline
        \diagbox{$\ell$}{$\ell'$} & $1\;2\;3$ & $2\;3\;1$ & $3\;1\;2$ \\ \hline
        $1\;2\;3$ & 0.052 & -0.018 & -0.206 \\
        $2\;3\;1$ & -0.018 & 2.218 & -0.242 \\
        $3\;1\;2$ & -0.206  & -0.242 & 0.973 \\
        \hline \hline
	\end{tabular}
	}\;\;
	\parbox{0.95\columnwidth}{\begin{tabular}{p{0.1\textwidth}<{\centering}p{0.1\textwidth}<{\centering}p{0.1\textwidth}<{\centering}p{0.1\textwidth}<{\centering}}
        \hline \hline
        \diagbox{$\ell$}{$\ell'$} & $0\;2\;3$ & $2\;3\;0$ & $3\;0\;2$ \\ \hline
        $0\;2\;3$ & 0.228 & -0.957 & 0.449 \\
        $2\;3\;0$ & -0.957 & 5.464 & -2.625 \\
        $3\;0\;2$ & 0.449  & -2.625 & 3.034 \\
        \hline \hline
	\end{tabular}
	}
	\label{tab:subchannels}
\end{table*}

The other signals can be analyzed along the same lines.
For identifying the microscopic process giving rise to the inelastic signal of frequency $4\omega_0$, the channels associated with permutations of the indices $0$, $1$ and $3$ need to be analyzed.
The process responsible for this signal is $\omega_0-\omega_1+\omega_3\to{\rm k}$. The corresponding signal photon wave vector fulfills $|\mathbf{k}_{\rm pw}|\approx 3.813\omega_0$ and is pointing at $(\varphi,\vartheta)\simeq(49.11^{\circ},78.93^{\circ})$.
This channel gives rise to $N(\mathcal{A}^{(4\omega_0)}|0,\infty)\simeq 129$
signal photons per shot. For the individual contributions constituting this number, see Tab.~ \ref{tab:subchannels} (b).
Finally, we turn to the two distinct signals with frequencies around $5\omega_0$.
These are triggered by the microscopic processes $-\omega_1+\omega_2+\omega_3\to{\rm k}$ and $-\omega_0+\omega_2+\omega_3\to{\rm k}$, respectively.
We summarize the detailed properties of these signals in Tab.~\ref{tab:channels}; this table also includes the parameters characterizing the $2\omega_0$ and $4\omega_0$ signals just discussed.
See Tabs.~\ref{tab:subchannels} (c) and \ref{tab:subchannels} (d) for the individual contributions constituting the signal photon numbers in these channels.

\begin{table*}[t]
	\centering
	\caption{Overview of the properties of the manifestly inelastic signal photon channels detailed in Sec.~\ref{sec:tracing}. For each signal photon frequency $\rm k\simeq n\omega_0$ with $n\in\{2,4,5\}$ we provide the longitude $\varphi$ and latitude $\vartheta$ characterizing the main emission direction as well as the number of signal photons per shot emitted into the solid angle $\mathcal{A}^{(n\omega_0)}$.}
	\begin{tabular}{cccccc}
	 \hline \hline
	 $\rm k$ & $2\omega_0$ & $4\omega_0$ & $5\omega_0$, A & $5\omega_0$, B \\
	 \hline
	 origin & $\ \omega_0-\omega_1+\omega_2\ $& $\ \omega_0-\omega_1+\omega_3\ $ & $\ -\omega_1+\omega_2+\omega_3\ $ & $\ -\omega_0+\omega_2+\omega_3\ $ \\
	 $\varphi$ & $319.11^{\circ}$ & $49.11^{\circ}$ & $23.41^{\circ}$ & $30.90^{\circ}$ \\
	 $\vartheta$ & $101.07^{\circ}$ & $78.93^{\circ}$ & $50.95^{\circ}$ & $32.36^{\circ}$ \\
	 $N(\mathcal{A}^{(n\omega_0)}|0,\infty)$ & $62.02$ & $129.40$ & $2.31$ & $0.46$  \\
	 \hline \hline
	\end{tabular}
	\label{tab:channels}
\end{table*}

\subsection{Implications of the channel analysis}

In the preceding section, we have worked out a strategy allowing us to trace all-optical signatures of quantum vacuum nonlinearity back to the underlying four-wave mixing processes and thus infer information about their microscopic origin.
To enable a clear measurement of a photonic signature of quantum vacuum nonlinearity it is desirable to maximize the signal at a given frequency and emission direction such as to achieve the best possible signal-to-background separation.
A complete assessment of the question which signal channel amounts to the most prospective one for an experimental verification, of course, requires one to account for many more details of a concrete experimental set up, including, e.g., the sensitivity and efficiency of the few photon detectors.

As the simultaneous measurement in several well-separated directions and at several frequencies is, however, highly unlikely with state-of-the-art technology, the typical challenge is to maximize the signal at a certain frequency and emission direction.
In this section, we sketch how the insights obtained in Sec.~\ref{sec:tracing} can be used to enhance a given signal photon channel.
Selecting a particularly promising signal, the channel analysis allows us to trace the microscopic origin of this signal and to modify the driving laser fields such as to enhance the signal in this channel, e.g., by redistributing the total available laser pulse energy into the individual beams.

This is especially obvious for the manifestly inelastic signals analyzed in detail in Sec.~\ref{sec:tracing}: while originating from the effective interaction of different subsets of beams, each of these four signals (cf. Tab.~\ref{tab:channels}) arises from the mixing of precisely three different driving laser fields.
Hence, in order to increase the signal photon yield in any of these channels individually, the driving laser beam which acts as a pure spectator can be switched off and its energy instead be redistributed into the other beams participating in the interaction.

Here, we illustrate this point using the example of a manifestly inelastic ${\rm k}\simeq4\omega_0$ signal originating in the microscopic process $\omega_0-\omega_1+\omega_3\to{\rm k}$, where $\omega_1=\omega_0$ and $\omega_3=4\omega_0$, respectively.
Obviously, only beams $0$, $1$ and $3$ are involved in this particular process.
Let us now remove the spectator beam $2$ and redistribute its energy into the other beams.
Our choice for the new beam energies is $\tilde{W}_0=\tilde{W}_1=\tilde{W}_3\approx 41.67\,{\rm J}$, maximizing the Fourier integral $\tilde{\mathcal{I}}_{013}\propto \sqrt{\tilde{W}_0\tilde{W}_1\tilde{W}_3}$; the other Fourier integrals do not support an inelastic channel or vanish for $\tilde{W}_2=0$.
The partition factors associated with this choice are $\tilde{q}_0=5/6$, $\tilde{q}_1=4/5$ and $\tilde{q}_2=1$, resulting in
$\tilde{W}^{\rm eff}  = \tilde{W}^{\rm loss} = 1/2\,W$; cf. Sec.~\ref{sec:amplitudes_losses}.
The new result for the number of signal photons in the manifestly inelastic $4\omega_0$ channel emitted into the angular area ${\cal A}^{(4\omega_0)}$ can straightforwardly be obtained from the corresponding signal photon number $N(\mathcal{A}^{(4\omega_0)})\simeq129$ determined in Sec.~\ref{sec:tracing}. It follows upon rescaling this number with an overall factor of  $\tilde{W}_0\tilde{W}_1\tilde{W}_3/(W_0W_1W_3)=36/25$, resulting in $\tilde{N}(\mathcal{A}^{(4\omega_0)}) = 36/25 N(\mathcal{A}^{(4\omega_0)})\simeq186$ signal photons per shot in this specific channel.

\section{Conclusions and Outlook}\label{sec:conc}

We have studied all-optical signatures of QED vacuum nonlinearity in the collision of several high-intensity laser beams differing in frequency, polarization and propagation direction. 
More specifically, we have focused on an example scenario envisioning the collision of four laser pulses, all originating from a single driving laser pulse, utilizing beam-splitting and sum-difference frequency generation techniques.
Such a scenario requires the availability of a high-intensity laser system of the multi-petawatt class, and thus, will become possible at various state-of-the-art and upcoming high-intensity laser facilities such as ELI-NP.
While we base our considerations on the availability of a single high-intensity laser of the ten petawatt class with specific parameters, our results can straightforwardly be rescaled to other laser parameters.

One of the goals of our study is to identify prospective signal channels allowing for an efficient signal-to-background separation.
To this end, we pay special attention to the question of how to efficiently infer information about the microscopic origin of prospective signatures of vacuum nonlinearity by means of a channel analysis.
This allows us to answer relevant questions, such as which laser beams participate in the formation of a given signal, and what is the specific interaction process inducing the latter.
In addition, we have explicitly demonstrated how this information can be used to enhance the signal photon number in a given signal photon channel.

For completeness, note however the difficulty of an absolutely background-free measurement in a real experimental set up.
Any practical imperfection such as a non-ideal vacuum in the vacuum chamber coming along with residual atoms and molecules in the interaction region may give rise to higher-harmonic backgrounds.
Still, these backgrounds can, in principle, be monitored (e.g. by rest-gas measurements, or geometric adjustments) and thus parametrically controlled to a large degree.
The full quantitative incorporation of such effects is outside the scope of the present idealized analysis.

We are confident that the concepts outlined and applied in the present study 
will prove very useful in future attempts at optimizing photonic signatures of 
quantum vacuum nonlinearity for given experimental parameters and constraints. 
Our formalism can also provide for an efficient basis to study recent 
alternative suggestions \cite{Ahmadiniaz:2020kpl,Sangal:2021qeg} for 
corresponding discovery experiments beyond the optical regime.

\acknowledgments

This work has been funded by the Deutsche Forschungsgemeinschaft (DFG) under Grant Nos. 416607684 and 416611371 within the Research Unit FOR2783/1.

\end{document}